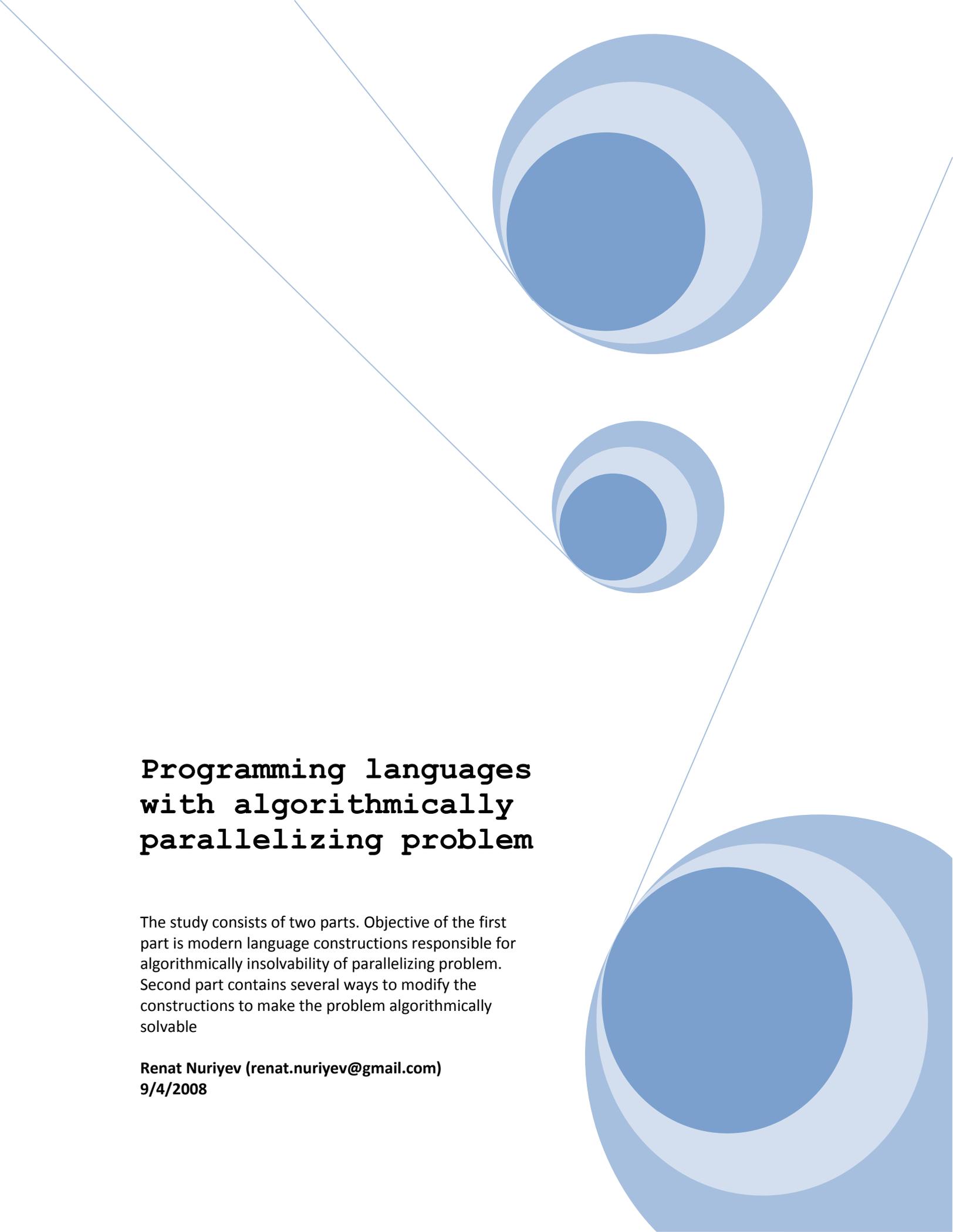

# Programming languages with algorithmically parallelizing problem

The study consists of two parts. Objective of the first part is modern language constructions responsible for algorithmically insolvability of parallelizing problem. Second part contains several ways to modify the constructions to make the problem algorithmically solvable

**Renat Nuriyev (renat.nuriyev@gmail.com)**
**9/4/2008**

Content

# Chapter 1.  Introduction

## 1.1.      Non-procedural and parallel programming

The multiprocessor systems take more and more areas of applications because of technological jump with chip integrations and resolution problems, their cost / productivity proportion and reliability.

The multiprocessor systems are used as final products as well as parts of them. Multi-core machines, PC-clusters, graphics processors, hard drive controllers are such examples. They are efficiently used in academic studies, industrial and military researches, in image processing, modeling of chemical and physical processes, databases searches. Now they became a main way to increase productivity of even home PCs.

The multiprocessor systems have different architectures and software because of high diversity of parallel and distributed processes in a real word. The main challenge here is not in standalone relatively simple software projects as in scientific modeling or image processing (we are not talking about science behind them) but in complex and big systems of business data processing. These applications are dynamic and have to accept new ideas and technologies.

To design such software with reasonable expenses of time, human resources and money we have to have software, technology and design patterns to support all stages of the projects oriented to run on a multi-processor systems.

We will focus on systems with large number of processors, say hundreds and more. For this class of applications it is practically impossible to create hundreds of threads without special tools and technologies.

Parallel/distributed/multithreading (PDM) programs are living in multi-dimensional time. Parallel software designer have to be like a good manager who is planning and orchestrating team (of processors) simultaneous working and communicating with each other. One can no longer think about program as "what steps I'd make if I do it myself".

Attractive approach is to use higher level of program languages - non-procedural. It is more specification oriented description. Execution details and in particular how to distribute data and calculations between processors are left to compilers.

In this paper we will consider an array of problems of creating a design environment for software for multiprocessor systems.

Let's made some classification of the parallel programming problems based on a life cycle of the parallel software.

Cycle goes through the same stages as sequential ones:

Stage 1. Identification of problems, ways of solving them, determining structure and hierarchy of sub problems.

Stage 2. Specification of input and output data for sub problems.

Stage 3. Definition of the set of tools and resources to develop the necessary tools.

Stage 4. Development and evaluation of algorithms for solving sub problems.

Stage 5. Determination of software design environment and coding the algorithms.

Stage 6. Local debug and test.

Stage 7. Global debug and test.

Stage 8. Documentation of the software, test, condition of usage.

Stage 9. Production use.

Stage 10. Maintenance.

In a parallel world each stage has its own appearances. Let's consider the stages one by one.

For the Stage 1 using multiprocessing system means better cost - efficiency in hardware. Alternative is to use network with equal cumulative power. Because of unnecessary duplication of peripherals its cost is much higher.

For the Stages 2-4 specification and programming for parallel processing solution better come with a higher level and more non procedural languages for specification and programming. Good examples are SQL, FORTRAN's vector oriented extension for scientific tasks.

Less procedural description means allowing more ways for program execution, including the parallel one. That is why a non--procedural programming is the important if not main stream for a parallel programming. Remarkable feature of non--procedural programs is their high compactness, they contain less execution details and hence shorter, easer to debug and maintain. It opens the way for full automation of business data processing and gain more than with a fragmentary automation.

Multiprocessor systems are a natural hardware for it.

Of course, the to-do-list for merging non--procedural and parallel languages is still big and includes some theoretical and system software problems, training and learning issues but the merger is here. In this paper we will consider formal base to build a non--procedural language for parallel processing of a large volume of data. Idea behind such languages is to use a system of formal logical definitions of data sets and specify the processing as transformations of data sets.

We will give formal definitions and describe studies of expression powers of a different type of formal definitions and complexity of the algorithms of building data sets.

For evaluation of practical usefulness we will describe a software project based on the data set definitions for business data processing and reporting.

We will also describe the torus multiprocessor systems with local two left and right neighbor communication. Main question is an efficiency of execution algorithms. We will show that locally defined behavior is not less efficient as centralized but has better tolerance to hardware failure.

Stage 4, the stage of design of algorithms, more than others depends on a communication topology, its speed and protocols. It is a known fact that algorithms designed for system with matrix connections (consisting of four neighbors nodes) is less efficient for the conveyor type system.

Stage 5 is a design/selection of software tools and programming languages. They vary a lot just because of diversity of ways and mechanisms of interactions in a real world processes. Probably one reason of a long life of a sequential program paradigm is that programmers think and describe in programs their own steps to solve the problem if they do it themselves. Programmer is a "single processor machine" and used to it. To program a multi-processor system he has to be a good manager, but not all people are ones.

Programming community accumulated a huge experience with sequential programming. It created an efficient Object-Oriented and message/event technologies and multithreading tools to build already not purely sequential programs.

For the stages 6, 7 of debugging and testing of parallel programs with explicit parallel control we need methods for detecting dead locks and estimating delays, we need theoretical research of existence and complexity of algorithms for it. Non--procedural languages are attractive for such problems because of smaller size, encapsulating parallel / multi-thread controls in compiler level. Non--procedural descriptions also work as high level specifications for themselves.

So detecting and parallel execution of program steps (not necessarily sequel) are important tasks for many reasons. Results of such studies could be used for selecting an architecture of parallel systems, more efficient executions and help to understand how to build languages for parallel programming. Good advantage of the task is a utilization of an existing programming environment: technologies, algorithms, experiences, training materials.

## 1.2. Levels of parallelizing

Parallel execution may be done on different levels:

- the inner-operator level when a parallel executable unit is part of operators like operations inside of arithmetic or logical expressions,

- the operator level, when a parallel executable unit is an individual operator,

- the level of blocks or segments of the program, where large pieces of code can be executed in parallel,

- the level of tasks when system has to run stream of independent tasks.

Each of these levels makes sense to use for appropriate tasks and parallel systems.

First and last levels are most commonly used. First level is used on hardware level such as conveyer processors and last one is used for execution of heavy atomic operations such as vector or matrix manipulations. All we need to do is to modify compiler for parallel execution of conveyer, vector or matrix operations. Another type of tasks that can be run such a way is a processing a sets of independent scenarios. Wide range of Monte-Carlo simulation based programs is good example.

Most attractive feature of these classes is that practically only few changes of the code have to be done.

Last level is already used in practice in corporate networking and Internet distributed tasks.

Success with the levels 2 and 3 depends on the complexity and the algorithmic solvability of the problem of detecting **information and logical dependency** of different steps of the program when result of one step used/overwritten with another one or used for selection of the operator to execute on another step. This problem is the key for all other considerations in this paper, so let's give it a short name – DILD.

Wide subclass of scientific programs can be automatically analyzed and converted for parallel executed (see full overview in [1]).

For this class of programs the parallel executions based on the fact that index expressions for them are linear. Let's call this class LI. In the chapter 2 we will show that if these expressions are more complex, say polynomials of the degree high than 3, then

DILD problem is **algorithmically unsolvable** even for total executable programs where for any path in program graph is an execution path for appropriate data.

Also we will show that for LI class the nested loops with n-1 dimension sets of independent steps cannot be linearly transform to the nest with even one independent loop. So it means that this method cannot be extended too far. But good news is that it proves the existence of parallelized, algorithmically full languages.

Stages 8-10 are less related to multiprocessor architecture and we will not consider them any deeper.

Parallelizing is useful not only for sequential programs. Even if program has some threads or processes there is still an open question if all potential parallel processes are used.

Two next remarks will determine farther studies.

First, especially important to run in parallel the most computationally intensive constructions as loops and recursions.

Second, the biggest area of computerization - processing of business data – is still mostly open for parallel programming.

## 1.3.        Model of program

We will use a general model of program as a set of functions and controls without details how they are constructed.

From this point of view Object oriented programs are a special type of control and special data structures which can be mapped to multi-dimension arrays with special data selection functions or a multi-base free algebra above of multi-dimensional arrays. Object is just a control for calling member functions or methods and for identification of object properties as elements of array with additional (objects) dimensions.

We consider algebra's operations as atomic ones and focus on parallel execution of sets of them. Function smart pointers and delegates are included in the model because on schematic level all uninterpreted functions and predicates are just smart pointers. Model does not cover parallel threads because of their dependency on state of operating system. So a program schema is a process of execution of some functions, predicates and their aggregations in subprograms or procedures. Data structures or XML data we consider as a multi-dimensional arrays with special functions for access of its elements.

Again, we are not going to use any special features of these functions except their signature.

### 1.4.    Notes to a history of the parallel processing

History of parallel processing is as long (or short) as a computer history itself. Because of a linguistic and informational barrier it make sense to separately mention studies in Russia.

Pioneer studies were published in [2]. The book discusses hardware and software problems, conveyer and independent organized calculation methods for many of well-known algorithms for scientific research. E. Evreinov also developed an idea of computer as a multi-dimensional net of modules "processors + multi-port communicator". These works were developed farther in researches of his followers: U. Horoshevskiy, N. Mirenkov, O. Badman, R. Nuriyev.

It is worth to mention the rarely citied fundamental researches of M. Zeitin related to collective behavior of finite automaton sets indirectly communicated through resources.

Very good notes about western world computer history is published [3]. It is hard to add something to it. We will just emphasise articles of Karp R. and Muller R. [4]. They made a necessary formalization and developed algorithm for parallelizing on operator level. Another pioneer work was published by Lamport L. [5]. It contains algorithm for detecting parallelism in Do loops in program with linear index expressions. Studies close to these was published by Badratinov M., Galiapin B., Nuriyev R. [6]. This article describes the program and its algorithms for detection independent and conveyer executable loops for the same class of FORTRAN programs.

### 1.5.    Parallelism as a partial order

Detecting if two steps are in DILD relation is an algorithmically unsolvable problem – there is no way to say in advance that some steps k and t are dependent. For program with simple variables only we are using more strong syntactically detectable relations such as if there is a variable in the left side of one operator and is an argument for another operator.

Execution order might be made weaker without changing the result. And execution with such order is a parallel execution. For any two orders $>_1$ and $>_2$ there are the top order $>_3$:

$$A >_3 b \Leftrightarrow (a >_1 b) \& (a >_2 b)$$

and low order $>_4$:

$$a >_4 b \Leftrightarrow (a >_1 b) | (a >_2 b)$$

which do not changes the execution result. So execution orders are an algebraic structure **full lattice** and hence we may talk about maximum parallel execution as a top element in this lattice.

It also gives us a good criteria for comparison classes of program with different controls. We may say that one class $C_1$ is more **flexible** than another class $C_2$ if the execution order in $C_2$ can be reproduced (constricted) in $C_1$, but not inverse.

## Chapter 2. Detection parallel steps in programs with arrays

### 2.1. Introduction

The problem of detecting of information and logically independent (DILD) steps in programs is a key for equivalent program transformations. Here we are considering the problem of independence of loop iterations – the concentration of massive data processing and hence the most challenge construction for parallelizing. We introduced a separated form of loops when loop's body is a sequence of procedures each of them are used array's elements selected in a previous procedure. We prove that any loop may be algorithmically represented in this form and number of such procedures is invariant. We show that for this form of loop the steps connections are determined with some integer equations and hence the independence problem is algorithmically unsolvable if index expressions are more complex than cubical. We suggest a modification of index semantics that made connection equations trivial and loops iterations can be executed in parallel.

We are considering not only algorithmic fullness of the programming language but also its data fullness when selection functions of elements of arrays are algorithmically full class too. These two features are independent. We suggest a modification of index semantics that made connection equations trivial and loops iterations can be executed in parallel. This modified language is a full in both senses.

We consider a DILD problem for programming languages with arrays and focusing on the syntax constructions determining the DILDs.

Transformations of programs, and parallelizing in particular, are based on fact that execution order for given start values can be changed without changing result. For program with single variables for fragment

x=f(u)   // step 1
y=g(x)   // step 2

we may say that step 2 informational depends on step 1.

But for the case of indexed variable

x[i] =f(u)   // step 3
y=g(x[j])   // step 4

information dependency takes a place only if values i and j are the same.

Let's consider two program fragments, structured and unstructured in Dijkstra's terms, when each loop has one entrance and one exit points as in figure 2.1.1:

Figure 2.1.1.

For the first fragment to identify an execution step for operator q we may use 2-dementional vector (p, k): p is a number of work out iterations of the upper loop and k is a number of lower loop iterations in the iteration p of upper loop. Step $q^{3,7}$ means that we are talking about such execution of the operator q when the lower loop made  7 iterations in a third iteration of the upper loop.

To identify a step for the second fragment we have to use variable length vector $q^{(i1,i2,\ldots)}$ meaning that we are talking about point of execution when upper loop executes i1 times, then lower executes i2 times, then upper executes i3 times and then inner executes i4 times and so on.

Good news is that any program can be algorithmically structured (Dijkstra's **goto** elimination problem) so that only nested loops may be used.

In section 2.3 we will extend the canonization result to recursion programs.

So we will consider from now and up only **structured programs:** programs in which each repetition steps are formed only with nested loops.

We will extend this idea by introducing **forced syntax principle** approach to modify of the language that syntactically implicit properties we need to solve problem in hands are encapsulated in a (new if necessary) syntactically explicit constructions.

Applying this principle to the DILD problem with indexed variables we will introduce a representation of a loop body as a chain of special subprograms (last one is called **kernel**, others are called **controllers**) which are responsible for calculation values of indexes for variables of next levels controllers or kernel but not for themselves. Such organized loops are called **separated** forms of loops. Other word, in separated loops the two loop body's functions of selecting data from some sets or arrays and processing that data are separated. For programs with separated loops we have enough syntactical objects to be able to define a condition if some set of iterations can be run in parallel.

A separated loop for processing connected list has two level controllers. First level controller selects node and the second one selects pointer to the next element.

In this chapter we'll prove that for FORTRAN like languages (or containing FORTRAN) the class of algorithmically parallelizing programs can't be remarkably bigger than the class with linear index expression. So to create language with algorithmically parallelizing programs we have to change some fundamental constructions of the languages and our goal is to point to these constructions.

## 2.2. Basic definitions

We will consider schemas of programs in which functions and predicates are symbols and they get interpretation (implementations) together with start data. But also we will allow some of them to have a fixed interpretation.

Let's A, X, F be finite alphabets without common elements and A*, X*, F* are a sets of strings in an appropriate alphabets.

**Def. 2.2.1.** We call a **memory M** a countable set of **cells** - two types of elements: **x** and $a[w_1, .., w_m]$, where **x** belong to the set X* and is called single cell, **a** belongs to A* and is called array of cells, $w_1, \ldots, w_n$ belong to X* and is called indexes of the cell.

A cell may get (be assigned to) a **value** from set $\Upsilon$ and keep it until next changing by a program operator. Cells without value are called **empty**. Each memory element has no more than one value and the value of element x will be denoted <x>, so <x>$\in \Upsilon$.

To keep this notation more close to programming languages let's suppose that one array name can't have different dimensions. Pairs of sets X, A and A, $\Upsilon$ do not have common elements.

**Def. 2.2.2. Variables** of a program schemas are elements of X* and expressions $a[K_1(z_1),\ldots, K_n(z_n)]$, where $a{\in}A^*$ is called array name, $K_1, \ldots, K_n{\in}F$ are called index functions, $z_1, \ldots, z_n$ are program variables from X*. Variables of a first type are called **simple** and of a second type – **indexed variables**.

For X={x,y}, A={a,b}, expressions x, y, xx, xyx are a simple variables, expression ba[xyx, yx] is an example of indexed variables, expression aa[ba[x]] is not a variable – expression in [ ] parenthesis has to contain only simple variables.

**Def. 2.2.3. Schema of program** is a 5D vector (**Opers, F, VarX, VarA, subP**) where **VarX** is a set of single variables, **VarA** is a set of array variables, **F** is a set of function an predicate symbols, **subP** is a set of (sub) schemas, also called **procedures**, **Opers** is a finite set of program instructions – strings of one of the forms:

   **a.** $\mathbf{l_1: x_0{=}g(x_1, \ldots, x_n)}$ **then** $\mathbf{l_2;}$ // assignment operator
   **b.** $\mathbf{l_1:}$ **if** $\mathbf{p(x_1,\ldots,x_n)}$ **then** $\mathbf{l2}$ **else** $\mathbf{l_3;}$// conditional operator
   **c.** $\mathbf{l_1:}$ **do** $\mathbf{P_1}$ **while** $\mathbf{p(x_1,\ldots,x_n)}$ **then** $\mathbf{l_2;}$// loop with body $P_1$ and iteration condition $p(x_1,\ldots,x_n)$,
   **d.** $\mathbf{l_1:}$ **do** $\mathbf{P_1}$ **then** $\mathbf{l_2;}$ // call sub schema or procedure $\mathbf{P_1}$.

Text after "//" is a comment, not a part of instructions.

Here $\mathbf{l_1, l_2, l_3}$ are called **labels**, $\mathbf{l_1}$ is called an **input label**, $\mathbf{l_2}$ and $\mathbf{l_3}$ are called **output labels**. $\mathbf{P_1}$ is called a **sub schema** and a set of it labels does not have common labels with upper level program or any other procedures, $\mathbf{p_1}$ is called a **repetition predicate**, $\mathbf{g \in F}$, $x_0, x_1, \ldots, x_n \in$ **VarX** $\cup$ **VarA**.

Output labels which are not input label are called **final,** and only one final label is allowed. Collection of labels of program or procedure P will be denote L(P).

One separate label $\mathbf{l_0}$ is called **start label** and its operator – **start operator**.
We assume that each of sets **Opers, F, VarX, VarA, subP** includes such sets for procedures too.

Schema is called **determined** if its instructions have different input labels.

**Interpretation** of the schema is a map of some finite set of memory elements to elements from $\Upsilon$ (they called **start value** for the program), function symbols interpreted as a maps $[\Upsilon^n \to \Upsilon]$, predicate symbols interpreted as a maps $[\Upsilon^n{\to}\{true, false\}]$.

Program variable **x** from **X** has a value <x> of the element of memory x, variable $\mathbf{a[K_1(x_1, \ldots, x_n),\ldots,K_n(x_1, \ldots, x_n)]}$ has a value $a[IK_1(<x_1>,..,<x_n>), \ldots, IK_n(<x_1>,\ldots, <x_n>)]$, $IK_j$ is an interpretation of $K_i$. Value of the variable is **empty** (not defined) if some of its index function has no value or one of memory element used as argument for index function is empty.

**The execution of operators** for a given interpretation consists of three parts - calculating some function or sub schema from the operator, changing the memory elements and marking some operator as next to execute.
More accurate, for operator of type:

**a**. calculate function Ig (interpretation of g) with $<x_1>$, …, $<x_n>$ arguments and put the result to cell $x_0$, mark next to execute operator with input label $\mathbf{l_2}$;

**b**. calculate $Ip(<x_1>,…,<x_n>)$, if it is true - mark next to execute operator with label $\mathbf{l_2}$, if it is false - mark next to execute operator with label $\mathbf{l_3}$;

**c**. calculate sub schema $P_1$ and then if $Ip(<x_1>, …, <x_n>)$ is true repeat this operator again, if it is false then mark next to execute operator with label $\mathbf{l_2}$;

**d**. calculate sub schema $P_1$ and then mark next to execute operator with label $\mathbf{l_2}$;

Operator execution is **defined** in following cases

**a.** all variables $x_1$, …, $x_n$ are defined and function $Ig(<x_1>, …, <x_n>)$ is defined as well;

**b.** all variables $x_1, …, x_n$ are defined and predicate $Ip(<x_1>, …, <x_n>)$ is also defined;

**c.** any iterations of $P_1$ are finished with final label and $Ip(<x_1>, …, <x_n>)$ after each iteration has a value **true** and becomes **false** after finite number of steps;

**d.** sub schema $P_1$ stops in his final label.

**Execution of schema** for a given interpretation is a chain of executions of marked operators starting with operator with start label $\mathbf{l_0}$.

Execution is called **ended successfully** if after finite number of steps some of marked label is final label.

Execution **ended without result** if some supposed to be calculated predicate or function does not have value or one of its arguments does not have value.

**Remark 2.3.1**. It is possible that schema will **run forever** without reaching final label. So schemas calculate **partially defined function**.

For shortness we will omit letter I in interpretation of functions and predicates if it is clear from context what it has to be - symbol or its interpretation.

## 2.3. Loops in a separated form

### 2.3.1. Introduction.

Canonic forms of the studying objects always have a theoretical and practical value because they allow classify objects, to work with smaller diversity and to use smaller description of the objects. We saw it for step enumeration of structured program. Also it nicely comes with level of abstraction for program design.

We hope that studying a canonic form for information processes in programs will help to simplify design, develop better debugging tools and increase the code reusability.

In this section we will show that any loop body can be represent as a sequence of sub procedures determine values of array indexes for next procedures, but not for itself or upper level controllers. Last sub procedure called loop's **kernel**, others are called **controllers** of corresponding levels. So we separate the loop's body execution in two functions – hierarchical selecting data and processing them.

Such implementation allows to reduce a debugging of complex processes of data selection in arbitrary arrays to the chain of debugging a simple blocks without arrays. It comes from functions of controllers. If each upper controller works properly, then data stream for this block works correctly and hence we need to check if this block is correct.

In C++ STL classes and generic collections in C# algorithms represent loops with fixed controllers for getting next elements from lists, maps, stacks and so on.

For theoretical studies this result is important because it gives syntactical objects for describing information connections in loops and, as it will be proved next, to divide the loops with different numbers of controllers or their connection graphs to unequal classes.

**Def. 2.3.1**. Let S(P) be a set of sub procedures of P, SS(P) be a transitive closure of this relations. Then sub procedure P is called **recursive** if P∈SS(P).

**Def .2.3.2.** Schema **P** is called **loop structured** (for short **L-schema**) if for **P** and for each of its sub schemas **P$_i$**:
  **a.**  a relation "input label is less than output labels" is a partial order, called "structural",
  **b.**  P$_i$ is not a recursive procedure.

So there is no recursion call, no spaghetti paths and no loops created by labels.

**Remark 2.3.1.** In L-schemas any repetition process can be generated only with nested loops.

**Remark 2.3.2.** L-schemas are structured in Dijkstra's terms: each loop has one start point and one exit point (we consider one final label only).

**Def. 2.3.3.** Two schemas with the same set of non interpreted symbols of functions and predicates are called **equal** if for each interpretation one schema finished successfully if and only if another schema finished successfully too and their state of memory is the same.

We can be proved the following
**Theorem 2.3.1.** There exists an algorithm to transform any schema to an equalent L-schema.

Full prove is mostly technical, too bulky and obvious. Instructions breach the label orders can be encapsulated in loops. Recursions can be replaced with couple loops and additional interpreted stack operations. Duplications of sub procedures can be eliminated by renaming and copying.

**Def. 2.3.4.** For a given instruction with input label **m** denote Ind(**m**), Arg(**m**), Val(**m**) sets of its index variables, set of it's argument variables, set of it's output variables. More accurately: if **m** is an instruction of type

a)      then Ind(**m**) is a set of indexes of array variables $x_0$, $x_1$, …, $x_n$; Arg(**m**) = $\{x_1, …, x_n\}$ ∪ Ind(**m**); Val(**m**)=$\{x_0\}$,

b)      then Ind(**m**) is a set of indexes of array variables from $\{x_1, …, x_n\}$; Arg(**m**)=$\{x_1, …, x_n\}$∪Ind(m);Val(m)=∅,

c)      then Ind(**m**)=∪Ind(k|k∈L(P)); Arg(m)=∪Arg(k|k∈L(P))∪Ind(**m**); Val(m)=∪Val(k|k∈L(P));

d)      then Ind(**m**)=∪Ind(k|k∈L(P)∪J$_1$, where J$_1$ is set of indexes of array variables of predicate p; Arg(**m**)=∪Arg(k|k∈L(P))∪Ind(**m**); Val(**m**)=∪Val(k|k∈L(P)),

For program or sub procedure P a set Ind(P)=$\cup$Ind(k|k∈P),
Arg(P)=$\cup$Arg(k|k∈P)$\cup$Ind(P), Val(P)=$\cup$Val(k|k∈P).

**Def. 2.3.5.** Loop **C** with non empty set of index variables is called **separated** if its body P is a set of instructions

    $m_0$: do $P_1$ then $m_1$;

    …..

    $m_n$: do $P_2$ then $m_{n+1}$,

where $\forall i \forall j$ [(Ind($P_i$)$\cap$Val($P_{i+j}$)=$\emptyset$) &(Ind($P_i$)$\cap$Val($P_{i-1}$)$\neq \emptyset$)], i, j∈N.

Here $P_1$,…$P_{k-1}$ are called **controllers** of levels 1,…,k-1, the last $P_k$ is called **kernel** of the loop.

**Def. 2.3.6.** The separated loop called **strictly separated** if
$\forall P_i$(i<k) Val($P_k$)$\cap$(Arg($P_i$)$\cup$Ind($P_i$))=$\emptyset$.

In other word, a kernel of a strictly separated loop does not change any variable used by any controller.

**Def. 2.3.7.** Schema is called **simple** if it consists only of instructions of type a) and d).

**Def. 2.3.8.** Two schemas are called **t-equally** if they are equal for each interpretation where functions and predicates are define for each arguments (also called **totally define**).

**Def .2.3.9.** Schema is called **forward oriented** if each loop has a body in which index of variable for any interpretation can be changed only before using. Syntactically, for each branch with array variables there is no operator changing its indexes with bigger label.

The following auxiliary statement can be proven.

**Lemma 2.3.1.** There exists an algorithm of transformation of any L-schema to the t-equal forward oriented schema.

**Idea of proof.** Let's have two instructions $S_1$ and $S_2$ in one branch of P. $S_1$ is executed early (input label of $S_1$ is less than for $S_2$) and uses index variable E and $S_2$ is executed later and changes E in one of the branches. Then we'll add new variable newE and add ahead of $S_1$ instruction

    newE = E;

and replace index E to newE in $S_1$.

Modified branch looks like next

….
newE=E;
x1=f(x[.., newE,..]…);
….
E=g(…);
…

It clearly equals to old branch for any interpretation of f and g and the branch now satisfies for forward orientation condition. By repeating such modification for any branch and loop body we will end up with forward oriented schema.

Now we are ready for main result of this section.

**Theorem 2.3.2.** There exists an algorithm to transform any loop C with arrays to t-equally separated loops with some n controllers and there is no equally separated loop with different number of controllers.

**Proof of the first part of theorem**. Let's B be a body of the loop C. According to lemma 2.3.1 we may assume that B has a "structural" partial order "$<$" on L(B). Let's for each branch collect such of instructions k with bigger input label than start instruction in order "$<$" and built a set of left side variables $Vs = \cup Val(k)$ until we meet instruction with indexes from Vs or get a final one. Let's call this instruction "limited", and continue with other branches. Process stops when all branches are visited and each ended with limited (red) or the final instruction. Visited set of instructions constitutes a first level controller.

To finish building the first controller let's add interpreted instructions with a new variable vLeb1 which will keep the output label of last executed instructions. It may looks like the following. Let $m_r$ is a final or limited output label. Then we'll replace $m_r$ with new label $m_{Aux}$ and add instruction

$m_{Aux}$: vLab1='$m_r$'.

We also add the next interpreted instructions to the rest set of instructions after removing organizer's instructions:

$m_{aux0}$: if (vLeb1==$m_1$) then $m_1$ else $m_{aux1}$ // start instruction of next procedure
$m_{aux1}$: if (vLeb1==$m_2$) then $m_2$ else $m_{aux2}$
$m_{aux2}$: ..,
here $m_{auxi}$ are new auxiliary labels.

These additions guarantee that after execution of the first controller the calculations will continue in an original order.

Clear, that if the rest set of instructions has instruction which changes indexes of others we may repeat the above process: mark as limited those instructions that have indexes from $\cup Ind(k|k \in L(P)) - Val(P_0)$ and separate next level controller $P_i$. If there are no such instructions then the loop has only i controllers and the set of rest instructions is a kernel.

Second part of the statement (about number of controllers) can be proven using special interpretations which we borrowed from Gödel's model theory. He developed it to study logical model (and called it model on constants). We will use this technique several times latter in this study.

Suppose we have some formal system with signature consisting of symbols of functions from **F** and predicates from **P**. Then the set $\Upsilon$ of values for standard interpretations is a set of terms T, built with elements of F and variable expressions. Formally T can be defined by induction:

**1.** Simple variables X, used in schema (finite set) are elements of T;

**2.** If a[r($x_1$,.., $x_{n1}$), ..] is an array variable in schema, then expression (string of symbols) '*a*[$r_1$($x_1$,…, $x_{n1}$), ..]' is in T for any $x_1$,…, $x_n$ from T.

**3.** term f($t_1$,…,$t_n$) for each $t_1$,..$t_n$ from T and **f ∈ F** also belongs to T.

**Standard interpretation of a functional symbol** f with n arguments is a map terms $t_1$,…, $t_n$ to the term (chain of symbols) 'f($t_1$,…, $t_n$)' if it is in T and is not define if it is not in T. In other word, the value of a function is a path of its calculation.

**Standard interpretation of a predicate symbol** is determined with finite set D, called a **diagram**, of expressions 'p($t_1$,…,$t_n$)' or '¬ p( $t_1$,…,$t_n$)' (only one of these two expressions is allowed to be in D ), where $t_1$, …,$t_n$ are from T.

Now we may prove the second part of the Theorem 2.2.2 that number of controllers is invariant for equal transformation.

Let's consider loop C for standard interpretation. By construction, each controller changes at least ones the indexes used in next organizer. So the value of indexes for standard interpretation have to be a word as '*a*[…, t,…]' where term t is a value of previous controller. Next level controller has to have a branch which changes indexes for next after it level and has to contain this expression. Otherwise if each path (for total schema it is also a branch) of such controller does not have this word in expression t for index, then all paths must be included in the previous levels.

Therefore execution of n controllers has to have value with n-1 deepness of parenthesis []. Hence two loops with different number of controllers can't be equal.

Proof is finished.

This technique may be used for more detailed classification of the loops. For example, from the proof it also follows that loops with different dependency graph between controllers also can't be equal.

### 2.4. Immediate information dependency between iterations

Each loop iteration has to depend from previous immediate iteration, otherwise iteration will be just a repetition of the previous iteration (memory used by iteration doesn't change) and hence, loop will run infinitely long. So iterations of whole loop body can't be executed in parallel, but parts of it can be. Iterations of level 1 controller have to have connections with body iterations; otherwise it can be run only one time.

For body with an indexed variable the result created on one iteration $n_0$ can be used on another iteration $n_1$ ($n_0$<$n_1$), not immediate next to $n_0$. To determine that the value of $a[K_0(i|n_0)]$ created on iteration $n_0$ used with indexed variable $a[K_1(j|n_1)]$ on iteration $n_1$ we have to solve an equation

$K_0(i|n_0) =K_1(j|n_1)$

for $n_0$ and $n_1$. The expression i|$n_0$ means value of i on iteration $n_0$.

For this equation we have to identify the system execution steps of the nested loops.

We'll use the following notations for nested loops and its elements. If $C_{i1}$ is a loop with body $B_{i1}$, then loops that belong to it will be denoted as $C_{i1,i2}$. In general $C_{i1,…,in}$ will denote loops belonging to the body of the loop $C_{i1,...,in-1}$. So depth of the nested loop is reflected in the index dimension of its name. The next diagram illustrates this notation:

Figure 2.4.1.

For simplicity we suppose that in a schema all instructions are different and each loop has a counter of iterations starting at 0 when loop instruction is initialized. Then for loop $C_{i1,...,in}$ a vector of counters $m=m_1,...,m_n$ of the inner loops $C_{i1,...,ip}$ for $p<n+1$ will be unique for the step of execution of any instruction q from its body and we will use the notation $q^m$.

### 2.4.1. Connection equation

Immediate connection between steps $q_1{}^{m1}$ and $q_2{}^{m2}$ (let $q_1{}^{m1}<q_2{}^{m2}$) takes a place when there is a simple or indexed variable value of it is created on the step $q1{}^{m1}$ and used on the step $q_2{}^{m2}$, or the result of $q1{}^{m1}$ is overwritten with $q_2{}^{m2}$ and it is the nearest step with such properties (there are no any operator $q_3$ and iteration $m_3$ ($m_3<m_2$) in between have a property like $q_2$).

For case of simple variable the nearest $m_2$ is the next iteration after vector $m_1$. In case of indexed variables $a[K_1(i)]∈Val(q_1)$ and $a[K_2(j)]∈Arg(q_2)$, immediate connection means that $K_1(i|m_1)=K_2(j|m_2)$ and there is no any instruction $q_3{}^{m3}$, where $q_1{}^{m1}<q_3{}^{m3}<q_2{}^{m2}$, with such a variable $a[K_3(p)]∈Val(q_3)$ that $K_3(p|m_3) = K_1(j|m_1)$ for $m_1<m_3<m_2$; $i|m_1$ means value for i calculated by some controller on iteration $m_1$.

So to detect information connections we have to have solution for equation:
**$K_1(i|m_1)=K_2(j|m_2)$.**

We will call it a **connection equation.**

It is a natural numbers equation which is a superposition of index expressions and functions of controllers on both sides. We can solve this problem for the system of linear equations but not for polynomials higher than 3 degrees.

It means that class of programs for solving system of linear equations, matrix operations and differential equations can be parallelized automatically. But this way is a dead end. Even for a case when connection equation is a polynomial, solving algorithm does not exist: we have a Diofant's equation problem.

**Remark 2.4.1.** We can show that the equation is the only problem – if we have solution for connection equation, no more algorithmically unsolvable problem is left for parallelizing.
This is technical result and we are not going to show it here.

### 2.4.2. Avoiding the connection equation

Let modify programming language (and call it language of **programs with predecessors**) by changing only semantics of index variables: instead of asking connection equation we may ask its solution. Now a variable with index expression like $a[g(i)]$ may be used only in right side of assignment operator and it means that on current iteration must be used value for $a$ that was created on previous iteration $g(i)$ and hence $g(i)<i$. Other word, index expression reflects a point in iteration space, not a point in an array.

There is no needs to show an index expression in any left side, it always the current iteration (when this operator was executed).

**Example 2.4.1.** Let's have a 4-point deferential approximation:

$f^k(i,j)=1/4(f^k(i-1, j)+f^k(i, j-1)+f^{k-1}( i+1, j)+f^{k-1}(i, j+1))$.

The equivalent program schema with predecessors is next system of loops

**for (k=1;k<N+1;k++)**
        **for (i=1;i<N+1;i++)**
                **for (j=1;j<N+1;j++)**
                        f = 1/4(f[k, i-1, j]+f[k, i, j-1]+f[k-1, i+1,j]+f[k-1, i, j+1]);

In reality last line has to be more complex because of start data. Start data or elements of input arrays might be introduced as negative number indexes. But for simplicity we are not showing calculation along the coordinate plains.

The natural way to execute a program with predecessors is to repeatedly run in parallel all iterations of loop's body fragments which have data. Obviously more efficient execution is to look out only iterations that just got data.

For example above controllers are simple and just increase indexes j, i, k. Their values do not depend on kernel (it a last line of code) and controller iterations can be executed before any kernel iterations. So we will get a list of 3D vectors (1,1,1), (1,1,2),…, (1,1,N),…, (1,2,1),…, (1.2,N), …, (1,N,N),…, (N,1,1),…, (N,N,N). For each of them we check if there is data for the arguments. In practice we have to check only points changed on a previous step.
At the beginning it is the point (1,1,1). Then data will be ready for 3 points (1,2,1), (1,1,2), (2,1,1).

Let P be a plain having these tree points, and **Norm** is its normal vector. Then at the next step data will be ready for iteration lying on next nearest plain with the same normal vector and including whole number coordinates. One can see that each next available for execution set of iterations is belonged to next parallel plain with integer points.

Each plain iterations on can be calculated in parallel. Really, for any integer point on the plain, all points used for this iteration are in an area strictly below this plain. It means that these points were already calculated and any point of plain can't be argument for another point of the same plain and hence points on one plain can be calculated in parallel.

So this 4-point approximation is highly parallel and for this 3D task parallel execution time is linear function of the dimension size.
Next figure 2.4.2 illustrates this case.

Figure 2.4.2.

Remarkable feature of this language is that it has a trivial algorithmically solvable problem of finding nearest iteration in which result of current iteration will be used. The index expression of right side variable is a solution of a connection equation. Still both languages are **algorithmically full**. It is easy to prove that there exists a fixed interpretation of functional and predicate symbols so that each algorithmic function can be calculated with a program from these classes.

There is no contradiction to Rise's theorem about insolvability of mass problems. The class of our programs is not an exact subclass – it is a full class.

For programs we have to consider two dimensions of fullness: algorithmically when any partial recursion function has a program for it calculation and **data fullness** when data selection functions are an algorithmically full class, any data structure can be represented.

Loops with simple controllers are very suited for parallelizing. They produced well organized data streams that can be efficiently implemented in parallel systems with separated memories. Questions are how useful such programs, how to organize parallel execution and how complex is it. The only comprehensive but expensive answer for that questions is a building of a real hardware and software system and applications of a wide enough area that executed in the system. But as it is shown in [7] there are many problems as well.

### 2.4.3. Cone of dependency and parallelism

Program with predecessors has a simple geometric interpretation. Let's consider hierarchy of loops with body B of the deepest loop. Then for each iteration $i_1,…,i_n$ we may calculate all immediate predecessors which values are used on this step. Repeating this for each predecessor, for predecessors of predecessors and so on we will get a set of iteration which we call a **cone of dependency**. It is clear that to get result for the current iteration we need results of any iteration from this cone.

Figure 2.4.2 represents the cone for iteration (k,i,j) for the 4-point example above:

Figure 2.4.2.

Any dependency cones for any point of any plain of iterations executable in parallel do not include each others, so they are independent from each other. This plain is a tangential to the cone at point (k, i, j) and the cone lies behind one side of it.

The relation between sets of independent and parallel executable sets of iterations is not simple.

Next example shows that **even if any set of iteration are lying on any line, parallel to some given one, the parallel execution for such sets does not exist**.

**Example 2.4.2.**

For (z=1; z<N; z++)
{
        For (x=1; x<N; x++)
        {

```
For (y=1; y<N; y++)
{
        for (p=1; ((x+p<n)&(y+p<n)); p++)
        {
                m_0: if (x>y) then m1 else m_2
                m_1:v=f1(u[p,x+p], u[x+p,p],v) then m_2
                m_2:v=f2(u[p,y+p], u[y,y+p],v) then m_3
                m_3: u[x,y]=f3(u[x,y], u[x-1,y-1],v) then m_s
        }
    }
  }
}
```

The cone of dependency for these nested loops is shown in figure the 2.4.4.

Here any set of iterations parallel to L (bold line) consists of independent iterations. But the two lines cannot be executed in parallel because dependency cone of one of them will contain points of another one. Thus any parallel to L line is a set of independent iterations, but only one line iterations can be executed in parallel.

**Figure 2.4.4.**

It can be proved that any forward loops can be transform to loops in separated form with the same technique as for ordinary program schemas. Number of its controllers and their connection topologies are an invariant for transformation. Therefore loops with different characteristics like these can't be equivalent.

Important feature of the forward loops is its **data selection fullness** - constructors may represent any functions. Hence any data structure might be represented and processed in parallel.

**Conclusion.** We've shown that problem of loop iterations independency contains the problem of solving connection equations. The last problem might be avoided by changing semantic of index expressions. The getting class is algorithmically full and has full data selections.

**Chapter 3. Processing stationary data stream**

**3.1. Introduction**

Loops with simple controllers are very suited for parallelizing. They produced well organized data streams that can be efficiently implemented in parallel systems with separated memories. Questions are how useful are such programs, how to organize parallel execution and how complex is it. The only full but expensive answer for that questions is a building a real hardware and software system and wide area applications that executed in the system.

Here we will consider processing business information – a main area of computer application. This area is one of the oldest in a human information processing – probably 4000 years old. Accounting clay tables were found in ancient Mesopotamia and Egypt. Data presentations and processing procedures are highly polished, so it makes sense to follow them.

Basic data units are rows, columns, tables and set of tables. More detail definition are in Chapter 4. Here we are focusing on control and organizing data streams for table processing on multi-processor hardware. Model of processing based on the practice when data comes to the system, are processed together with some distributed data or waits for them and are destroyed at some point.

The best feature of tables are that their data accompany with its semantic: table tags such as type (telephone bill, office expenses), table names - date, department name, row names, column names allow to recreate the semantic of the data and to verify them, to design algorithm of processing, to communicate and to make decisions.

Semantically finished results appear after processing whole unites or combination of units: after processing each row, after processing whole table alone or together with row of another table, after processing one table with another table. Intermediate result does not have too much sense itself. Often only some set of tables have a real value. Set of reports without report from one department rarely have value for top company management. In most cases a missing pieces of some tables are an exceptions to be reported.

For processing several tables people use priorities to look out: take a row in one table and look up rows of another table while some rows would be found and so on.

The point is: in a table processing information relations between steps are regular, predictable, and thanks for it are efficient for parallel execution. More detailed data presentation as a system of tables (similar to XML) we will consider later when describe the real system for office data model.

Further we will focus on parallel processing of system of tables as a system of loops with a regular data flow.

**Def 3.1**. Let $\{L_1, \ldots, L_n\}$ and $\{K_1, \ldots K_m\}$ are a two sets of lists called input and output lists. Let lists have priorities for their look out - natural numbers. And let each list have a pointer $p_j$ of the position in the list with priority j. So these pointers are a n-dimensional vector with natural numbers components, let's call it farther a **vector-pointer**.

Loop, calling **priority loop,** on each step is running two procedures. First is moving vector-pointer to next one in lexicographical*[)] order (we will call it LO-order for short). Second represents the function which is taking values corresponding to vector-pointer as arguments and putting results to positions pointed in output lists. So a corresponding program schema doesn't have nonlocal single variables.

Example of lexicographical ordered n-dimensional vectors are the counters of the of loops:

```
For (i₁=0; i₁<N₁; i₁++)
        For (i₂=0; i₂<N2; i₂++)
                ……………..
                For (iₚ=0; iₚ<Nₚ; iₚ++)
                {
                ……
                }
```

Next example shows that these very restricted programs are doing all jobs for reports generator.

**Example 3.1.** We want to design service program to control limit for office expenses for company's divisions. It uses data in tables OfficeExpenceLimitation and PricesOfOfficeSupplies, it starts when table of type OfficeSuppliesBought is entered or updated.

Design Studio ask name of new application (say Expenses) and ask to fill up the system table named Program Initiation Table. For our task input is shown below in red:

-----------------------------------------------------------------------------------------------------

*) Vector $v_1=(v_1,…,v_n)$ is lexicographically greater than $v_2=(w_1,…,w_m)$ if m>n or m=n and $v_1>w_1$ or $(v_1=w_1)\&(v_2>w_2)$ and so on.

Type **Program Initiation Table**, name **Expenses**

| Input | : output | :allow to delete |
|---|---|---|
| 2.  OfficeExpenceLimitation | : 1.Office Expence Limitation  : | |
| 3.  *OfficeSuppliesBought | : | : * |
| 4.  PricesOfOfficeSupplies | : | : |

So this table contains input and output tables types, their priorities (number in ahead of table) and processing activated table type (marked with * ahead). Another mark * in last column means that this table do not needed to after this processing.

Next Design Studio shows one by one the empty tables from previous **Program Initiation Table** and ask to input short names for columns which be used in the program:

Type **OfficeExpenceLimitation**

| Department | : Limit amount | : Out of limit |
|---|---|---|
| DepL | : Limit | : LimitMark |

Type **OfficeSuppliesBought**

| Department | : Supplier type | : Quantity |
|---|---|---|
| DepB | : Sup | : Quant |

**Type PricesOfOfficeSupplies**

| Product | : Price | |
|---|---|---|
| Prod | : price | |

Final is a table for processing rule

**Type PROGRAM**

| Conditions | : action | :result | : comments |
|---|---|---|---|
| **C**: (DepL=DepB)& | : Limit-Quant*price | : Limit | : |
| &(Prod=Sup) | : | : | : |
| | : | : | : |
| **O**: Limit<0 | : * | : LimitMark | : exceeded limit |

----------------------------------- :

Here in column "Conditions" it is allowed to use three types of labels for functionally different logical expressions.

Label V means that processing is allowed, false means "should not start this process".

Label C means that current combination of input rows allowed to process. False means stop processing this combination of rows and take another one.

Label O means that instructions in section starting from this line up to next horizontal line has to be executed, false means ignore instruction from this section. Section for O condition might be nested.

Now each time when a new table of type **Office Supplies Bought** will be entered/edited, this program will be initialized automatically. If any of the other tables mentioned in first column of **Program Initiation Table** is absent then process will be postponed until their appearance.

Clear that because of priority of the only output table **Office Expense Limitation** is not smaller than 1, each raw of table **Office Supplies Bought** with priority 1 can be processed in parallel.

Mentioned here Design Studio is a part of operational environment GNOM developed for Russian Census Department.

### 3.2. Parallel execution of priority loops

**Introduction**

In this paragraph we'll establish syntactical criteria for parallel execution of priority loops based only on signature of functions.

Also we'll consider an architect of the multi-processor systems for efficient execution the priority loops.

### 3.2.1. Syntax criteria for parallel execution of priority loops

**Def. 3.1.1**. Let P be a program schema. **Right labeled path** in P is a sequence of its instructions where each non start input label is an output label of the previous instruction. Program schema is called **total** if each its path is right labeled and there is an interpretation where this path is an execution path.

Mostly we will consider priority loops as a **nested ones with only one loop in bodies** and call them **priority nested loops**. Each loop starts with new lookout of table starts, and finished when the end of this table be reached. The body B has only local variable – values of one iteration is not available for others. All connections between one loop iterations go throw upper priority lists.

Next picture shows the graph of priority nested loops:

**Figure 3.1.1.**

Priority loops are **one time written memory schemas** – elements of any output list with priority $r$ can be created on one iteration of level $r$ loop and never be changed later.

**Theorem 3.1.1.** Let $C$ to be a total priority nested loops with a kernel B. Let S be a set of lists from $\text{Arg(B)} \cap \text{Val(B)}$. Then if priority of any list from S is more than $r$ then iterations of a loop $C_r$ (or a body B for $r=1$) are independent.

**Proof.** Let list $a$ with priority r to be in S. Then a value of $a[j]$ for any $j$ is available for change and use in any loop $C_d$ for d<r. And hence these iterations are dependant throw $a$. If d is a minimal with such list $a$ then iterations of any loop $C_e$ (e<d and e>0) are independent because variable of B are local and any of output list's cells are new for each iterations. End of proof.

Priority loops can be generalized for the case when bodies of loops $C_i$ are a sequences of three procedures $H_i C_{i+1} K_i$. Hi is called a starter, Ki is called a final for lists with priority i. $H_i$ processes the first, $K_i$ - the last element of lists of priority i, so they are analogs of OOP constructors and destructor list objects.

Graphic representation of extending priority loops is next

Figure 3.1.2.

Upper oriented arrows are input lists and down oriented are output lists.

### 3.1.2. Algorithm for parallel execution of priority loop on a multi-processor system with a ring topology

Ring system is a set of computers with local connections to left and right neighbors. The biggest advantage of this topology is its similarity and ability for independent communication for different pairs to avoid traffic.

Because of similarity it makes sense to equally distribute all lists or if list is small – distribute whole copy of it. Then execution process just follows to this data distribution.

Let's describe execution for priority loop with 4 input lists A,B,C,D, with independent second loop and 4 modules M1-M4 multiprocessor system. We suppose that list B is independent and has 16 elements. Next **parallel execution diagram** shows what data are processing in each module in time 1,…, 6:

| Time/module | 1 | 2 | 3 | 4 | 5 | 6 |
|---|---|---|---|---|---|---|
| M1 | 1,1,C, D | 1,2,C,D | 1,3,C,D | 1,4,C,D | 2,1,C,D | 2,2,C,D |
| M2 | 1,5, C, D | 1,6,C,D | 1,7,C,D | 1,8,C,D | 2,5,C,D | 2,6,C,D |
| M3 | 1,9,C,D | 1,10,C,D | 1,11,C,D | 1,12,C,D | 2,9,C,D | 2,10,C,D |
| M4 | 1,13,C,D | 1,14,C,D | 1,15,C,D | 1,16,C,D | 2,13,C,D | 2,14,C,D |

Here module M1 at time 4 is processing element 1 of list A, with element 4 of list B with each pairs of elements of C and D.

It is easy to interpolate this example to general cases.

Execution can be done if modules shared memory. But for local memory we need to do additional job for moving data between modules.

Let's consider the case when all lists A,B, C (and no D for short) are distributed. Then parallel processing became a little slower (here C is divided to 4 pieces C1, C2, C3, C4 and they distributed in M1, M2, M3, M4 correspondently):

| Time/module | 1 | 2 | 3 | 4 | 5 | 6 |
|---|---|---|---|---|---|---|
| M1 | 1,1,C1 | 1,1,C2 | 1,1,C3 | 1,1,C4 | 1,2,C1 | 1,2,C2 |
| M2 | (C2) | 1,6, C1 | 1,6,C2 | 1,6,C3 | 1,6,C4 | 2,6,C1 |
| M3 | (C3) | (C2) | 1,9,C1 | 1,9,C2 | 1,9,C3 | 1,9,C4 |
| M4 | (C4) | (C3) | (C2) | 1,13,C1 | 1,13,C2 | 1,13,C3 |

Now each time after "hand shaking", each module sends its C fragment to the left and gets new piece of C from right neighbor. Module starts process when the right piece has arrived. For example, module M3 at time 1 sends his fragment C3 to M4 and gets C2 from M2, at time 2 it sends C2 to M4 and gets C1. The task is finished when all pieces are back to each module.

### 3.3. Local control for the priority loop execution

Above execution requires some additional jobs:
-   equally distribute the list fragments,

- exchange fragments,
- sorting these fragments,
- detection defected module (out of order, dead lock)

We will solve these tasks under next assumptions
- each module may communicate with 2 fixed neighbors, only one in a time and after communication with one neighbor(say right) next it communicates with another (now left) neighbor,
- communications are occurred asynchronously (no external synchronization signal),
- neighbor might be busy when another tries to connect and it has to wait,
- communication speed is as good as reading from memory
- number of modules is even and modules are numerated 1,…, 2n .

Farther we will show that all above tasks might be solved with decentralized algorithms when control of each module based on his own and next neighbors states. States are controlled with some finite automates, it is light, no needs in too much resources.

Main challenge comes from fact that waiting time may linearly depends of the number of modules. Let procedure T has two paths with different time of execution, say 1 and 10 secs. Then if execution time for each step k>1 and module k is 10 secs, and 1 sec for others, then waiting time for module k+1 to get response from module k+1 will be 9*k secs.

To make a waiting time constant or at least logarithmically dependent from number of modules time to time we may run next automat in each module.

All algorithms are based on odd-even modules pair communications.

At the start time all odd modules are in state **q** – to send a package to the left neighbor, all even modules are in a state **p** – get a package from right neighbor. After finishing communication, even with empty packages, modules change their states, now odd modules are waiting packages from right modules. Then next graph is its state transition graph:

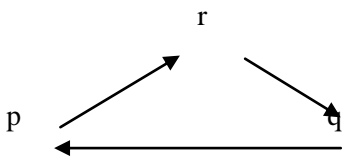

An almost-synchronizing algorithm is presented with next automate with two input symbols K and F. The input symbol K means getting package, F – getting synchronization flag. Commands of the automate are expressions of the form

$(q_1, S, m) \rightarrow (D, q_2)$,

where $q_1$, $q_2$ are a states, S is an input symbol, m - a finite memory symbol. Command means: if automate is in a state $q_1$ and got symbol S to its input and memory value is m, then do action D and then change state to $q_2$. Symbols p, r, q are states where automate does the same as above; in state f automate generates flag F with module's ID number.

In situation (r,F,i) automate is checking if a flag F carries module's ID. If ID is not his, automate waiting for getting another flag, if ID is his – gets a package from right neighbor.

Next diagram is a state transition graph

Figure 3.2.1.

Corresponding automate is doing "handshaking" with his left and right neighbors even if nothing to send to guarantee that waiting time wouldn't grow too much.

The comparison of the states of neighbor's automates gives a clue for locally detecting of an incorrectly functioning neighbor.

### 3.2.1. Decentralized control for parallel executions of a priority loop

Here we will evaluate execution time of the algorithms of solving two tasks mentioned at the beginning of this paragraph.

**Task 1: equally distribution a list.** One solution is to collect sum of numbers of elements in each module, find average, send it each module and let to send to the left neighbor all elements above this number. Complexity of the algorithm is >3n. Worst case scenario is when elements will travel a whole ring and at some point all will be gathered in one module.

Another strategy might be to keep doing equal parts for pairs of neighbors: numerates modules with numbers 1,…, n, divides system to odd-even neighbor pairs (1,2), (3,4), (5,6), equalize the number elements (if the joint number is odd, put bigger to left module) in each such pair, then separate the pair of neighbor modules from different previous pairs (n,1), (2,3), (4,5),.... Again equalizes them and make pair (1,2), (3,4) … again and so on.  It can be proved that at most after 2n+1 step the task will be solved.

**Example**. Let's have a ring with 10 modules with 4, 56, 34, 10, 20, 50, 12, 73, 16, 23 elements each. Next diagram shows how this local distributed algorithm will work:

```
Time 1:     ( 4, 56),  (34, 10), (20, 50), (12, 73), (16, 23)
      2:      30), (30, 22), (22, 35), (35, 43), (42, 20), (19,
      3:     (25,  26), (26, 29), (28,  39), (39, 31), (31, 24)
      4:      26), (25, 28), (27, 34),  (33,  35), (35, 28), (27,
      5:     (27, 27), (26, 31),  (30   34), (34,  32), (31, 26)
      6:      27), (27, 29),(28,  32),  (32, 33), (33,  29),  (28,
      7:     (28,  28), (28, 30), (30, 33),  (32,  31), (31, 27)
      8:      28), (28, 29), (29,  32), (31, 32), (31, 29), (29,
      9:     (29, 29), (28, 31), (30,  32), (31,  30), (30, 29)
     10:      29), (29, 30), (29, 31), (31, 31), (30,  30), (29,
     11:     (29, 30), (29, 30),  (30, 31), (31, 30),  (30, 29)
     12:      29), (30, 30), (29,  31),(30, 31), (30, 30), (29,
     13:     (29,  30), (30, 30), (30, 31), (30, 30),  (30, 29)
     14:      30), (29, 30), (30, 31), (30, 30),  (30,  30), (29,
     15:     (30,  30), (29, 30), (31, 30), (30, 30),  (30, 29)
     16:     30),  (30,  30), (29, 31), (30, 30), (30,  30), 29
     17:     (30,  30),  (30,  30), (30, 30), (30, 30), (30, 29)
```

For odd case it works 2n+1 steps, where n is the nearest even number.

**Task 2: sorting a list.** Sorting process starts with the equalization. Then each module locally sorts its part of list and merges lists in odd-even pairing and divides it in two equal parts. It can be proved (proof too long to put it here) that in n steps system will get the sorted list.

## Chapter 4. Non procedural language for parallel programs

### 4.1. Introduction

In a chapter 2 we've noted that massive data processing in modern languages can be organized with multi-level controller loops. In a previous chapter 3 we saw that for a wide area of applications massive parallel processing studio might be built around simple stream of data and cover a big application area. Modern popularity of XML data presentations means that loops with multi-level controllers and hence a complex data streams comes in practices too.

Probably building non- procedural languages is the most prospective way for parallel programming just because non procedural means no fixed way for execution: we are free to select what is better fits the case. Latter we will see that finding a way of execution for a given nonprocedural description might be a heavy task too.

Parallel programs may be considered as a process of transformation some data sets on each step. And such staff is already in practice. For example, airline company decision "ticket price for passengers travel more than 5 000 miles per year have to be 90% of original price" is based on definition and transformation of set of tickets of passengers with some property.
A sequel program is just a case when these sets have of one element.

There are two major ways for definition a set of data: procedural and non-procedural. Procedural way means we may define some algorithm for selection data to the set. Non-procedural way is to define a set without description how to create it.

Rational data base languages are an example of the last way. But the languages are not the most powerful. Next definition is an inquiry that can't be expressed in the languages: **select a set of people sympathetic to each other**. This definition can't be used in a database - it requires a name of the data set. It also does not allowed in mathematics because of next "good" set paradox.

Example of nontraditional inquiry might be next: **select a set of people sympathetic to each other.** This definition can't be used in database. It requires a name of the set. It is also not allowed in mathematics because of next "nice" set paradox.

Set X is called good if ¬X∈X. If U is a set of all "nice" sets is U good or not? Both possibilities are ended with contradictions.

If U is "nice", then by definition ¬(U∈U) is true. But U contains all "nice" sets and therefore it must include U itself and U∈U is true. We got a contradiction to our assumption.

If U is not "nice" set, then U∈U is true. But U consists of such sets X that ¬(X∈X), so ¬(U∈U) should be true for U. Contradiction.

**Def . 4.1.1.** We will call a **date** a pairs **(d, v)**; **d** will be called an **attribute** and **v** will be called a **value**. Elements **d** and **v** are taken from some countable sets D and V. They may be equal dimension Cartesian product too, then $d=d_1,\ldots,d_n$ and $v=v_1,\ldots,v_n$.

Definitions of data set used here are constructed hierarchically from elementary definitions taken from mathematics and computer science practice. We recognize four types of them:

- **enumeration**, when we point an elements of the set;

- pointing **properties** of set's elements in relation to the elements of the same or another sets;

- using **induction** when we fix the base set of elements and induction predicate **p** (rule) by which we are saying that if an element is in this relation with already established elements, then this element belongs to the set too;

- defining a set of elements as a **functional image** of elements of another sets.

Each definition also determines the name of the set or names of sets if we are using parameterized naming. So to define a single set we generally have to use a system of definitions with some sets of names.

**Def . 4.1.2.** Sub systems of definitions are using only sets defined in the same system are called *closed* (and *open* otherwise). Set names defined in the system also is called *internal*, all others – *external*.

To be a fully defined, system with external names has to have, besides of start data and interpretation functional and predicate symbols, some start or *a priory* family of sets. Then a system of definitions is a function of *a priory* family of sets to the family of sets defined by this system.

This view gives us a way to compare different subclasses of definitions by comparing classes of associated operators.

**Def . 4.1.3.** We may say that a **class of definition is more powerful than another** if class of associated operators includes the class of operators associated with another subclass of definitions.

We will show that no one of types definitions can't be eliminated without decreasing expression power. Also extending signature with interpreted predicate

$$p(x,S) \equiv (x \in \mathcal{S})$$

increases the expression power.

In paragraph 4.6 we will consider an algorithmic complexity of building sets satisfying to system of the definitions from different classes.

### 4.2. Subclasses of sets definitions

**Def . 4.2.1.** Let N to be a set of natural numbers. **Alphabet** for system of definition is a union of punctuation symbols, parenthesis, special symbols $\forall, \exists, \equiv, \in, \cup, \cap, +, -$ and non-intersected sets P, C, S, F, V, where

$P = \{p_j^i | i \in N, j \in N\}$

$C = \{c_i | i \in N\}$

$S = \{S^i | i \in N\}$

$F = \{f_j^i | i \in N\}$

$V = \{x_i, y_i, z_i, u_i, v_i | i \in N\}$.

Elements of P are called **symbols of predicates**, elements of C are called a **symbols of constants**, elements of F – **symbols of functions**, S – **types of set names**, V –**variables.**

**Def. 4.2.2.** Set of **terms** T is defined by induction:
a. empty term $\lambda$ is a term,
b. elements of V and C are terms,
c. if $t_1, \ldots, t_k$ – are terms, $f_j^k \in F$, then $f_j^k(t_1, \ldots, t_k)$ is a term.

Let's for term **t** denote v(t) the deep set of variables of t from V. More formal v(t) is a minimal set with next 3 properties:
a. if $t \in V$, then v(t)=t;
b. if $t = f(t_1, \ldots, t_k)$, then $v(t) = v(t_1) \cup \ldots \cup v(t_k)$;
c. if $t \in \{\lambda\} \cup C$, then $v(t) = \emptyset$.

**Def. 4.2.3.** An element of $S \times T \times N$ is called **schema of set names** and is denoted $S_j^t$ ($t \in T$, $i \in N$ and is called a type of name).

**Def. 4.2.4.** Expressions $x \in S_j^i$ or $S_j^i \in S_y^m$, where $x \in V$, $y \in T$, are called **elementary selectors** of variable x or of set names $S_y^j$ of type j.

**Def. 4.2.5.** A finite set U of elementary selectors $x_1 \in S^{i1}_{t1}, \ldots, x_n \in S^{in}_{tn}$ with all different $x_1, \ldots, x_n$ is called a **selector level** if $(v(t_1) \cup \ldots \cup v(t_n)) \cap \{x_1, \ldots, x_n\} = 0$. The set $(v(t_1) \cup \ldots \cup v(t_n))$ is denoted $U^-$, $\{x_1, \ldots, x_n\}$ is denoted $U^+$.

**Def. 4.2.6. Hierarchy of selectors** U is called a such finite set of selectors levels $U_1$, $\ldots U_n$, that for each $x \in U_i^-$ also true $x \in \cup^k_{(i<j)} U_j^+$ and $\forall i \forall j ( U_i^+ \cap U_j^+ = \emptyset)$.

A hierarchy of selectors is called **closed** if $U_n = \emptyset$.

Other word, a hierarchy of selectors defines domain for each variable of the form.

**Def. 4.2.7. Elementary form of subset data definitions** (**form of definition** for short) is called one of the next expressions:

α. $S_c^i = \{c_1, \ldots, c_k\}$,

β. $S_t^i = \{Q_1 x_1 \in S_{t1} \ldots Q_n x_n \in S_{tn} \, \mathbf{p}(x_1, \ldots, x_n, z_1, \ldots z_m); U\}$,

γ. $S_t^i = \{S_{t0}^{i0}, Q_1 x_1 \in S_{t1} \ldots Q_n x_n \in S_{tn} p(x, y, x_1, \ldots, x_n, z_1, \ldots, z_m); U\}$,

δ. $S_t^i = \{t_0 (z_1, \ldots, z_m), U\}$,

where $S_t^i$, $S_c^i \in S \times T$, $Q_1, \ldots Q_n \in \{\forall, \exists\}$, $p \in \mathbf{P}$, $\{z_1, \ldots, z_m\} \subset v(t)$, U is a closed hierarchical selector for variables from $v(t) \cup \{z_1, \ldots, z_m\}$; $c_1, \ldots, c_k \subset C \cup \{S \times C\}$, $x_1, \ldots, x_n, z_1, \ldots, z_m \in W\{S \times W\}$, $t \in T \cup \{S \times T\}$, $v(t) = \cup U_i^+$.

Variables from $v(t)$ are called **parameters** for this form.

Idea behind of forms is next.

Form of type α defines set with name $S_c^i$ and elements $c_1, \ldots, c_k$.

Form of type β defines family of sets when variables of the term t (names) are taken from sets in selector U and each set is defined by relation **p** its elements to elements of the same or another sets $S_{t1}, \ldots, S_{tn}$ expressed with the first order theory formula

$Q_1 x_1 \ldots Q_n x_n \, \mathbf{p}(x_1, \ldots, x_n, z_1, \ldots, z_m)$

and $x_1, \ldots, x_n$ are taken from sets with names $S_{t1, \ldots,} S_{tn}$ correspondingly.

**Example of form** $\boldsymbol{\beta}$. List of products Pr is an example of set consisting of single elements and sets – product may include of another products. The form

$S^{india} = \{\forall S_x \in S^{india} \; \forall y \in S_x \, p(y)\}$,

p(y) means product **y** is produced in India, defines the set of products completely produced in India.

Form of type γ defines sets with names $S_t^i$ when variables of term t are selected with selector U. Each set $S_t^i$ is defined by induction and $S_{t0}^{i0}$ is a base of induction, formula $Q_1 x_1 \in S_{t1} \ldots Q_n x_n \in S_{tn} p(x, y, x_1, \ldots, x_n, z_1, \ldots, z_m)$ is an induction step – if $y \in S_t^i$ then any element x from universe also belong to $S_t^i$ if this formula is hold.

Let $G$ be a graph $(V, R)$, $V$ is vertices, $R$ is a set of edges. The form

$S = \{S_0, p(x, y), y \in V\}$,

where $p(x, y)$ is interpreted as edge $(x, y) \in R$, define a minimum connected components of the graph $G$ which includes vertices of $S_0$.

Form of type δ defines, probably better to say creates, a "secondary" data. This form determines sets $S_{vt}^i$ where vt is a value of term **t** when its variables selected with selector U. Elements of set $S_{vt}^i$ are values of the term $\mathbf{t_0}$ with arguments selected with selector U. It may be elements out of start data. After applying the definition new data are included in data universe.

Generally, forms are based on another definition forms. So to define a set we have to talk about a system of definitions.

**Def. 4.2.8.** A finite set **F** of forms are called a **system of forms** if:

a. left parts of forms are different types;

b. for each set name S type of j the system **F** has a definition of that type.

### 4.3. Interpretations of systems of set definitions

**Def. 4.3.1.** Set of symbols of predicates, functions, variables and constants used in forms of a system F is called a **full signature** of **F** and **denoted $\sigma_F$**.

**Def. 4.3.2.** Let $V = D \times V$ is a **data universe**. To define an **interpretation J** for system of set definitions **F** means to define **start data $\Omega$** – a finite set of data, finite set of secondary data $\Omega^\lambda \subset \cup f_i^k[V^k]$, and map

- each **symbol of constant** to element of $\Omega \cup \Omega^\lambda$;
- each **predicate symbol** $p_i^j \in \sigma_F$ to map $[V^j \to \{true, false\}]$
- each **functional symbol** $f_i^j \in \sigma_F$ to map $[V^j \to V]$.

**Terms** used in a form are substitutions of function symbols and variables.

For a given interpretation **J,** set names $S(\mathbf{F}, \mathbf{J})$ also get a concrete names:
$S^{conc}(F, J) = \cup\{S_v^i | v = t(v_1, \ldots, v_n), \{v_1, \ldots, v_n\} \subset \Omega \cup \Omega^\lambda; U\}$, where $S_v^i$ are any schema of names in F and U is a selecrors for variables of $\{v_1, \ldots, v_n\}$.

For a given system F and interpretation J we will also use expressions $\Omega = \Omega(J, F)$, $\Omega^\lambda = \Omega^\lambda(J, F)$, $D = D(J, F)$.

### 4.4. Named data subsets defined with a system of definitions

Let J be an interpretation of a system F, **$M$** is a set of pairs $(S_v^i, M_v^i)$, where $S_v^i \in S(F,J)$, $M_v^i \subset \Omega \cup \Omega^\lambda \cup S(F, J)$.

**Def. 4.4.1.** Let's a set of names from pairs of **$M$** denote as $S(M)$, a data subset with name $S_v^i$, denote as $M(S_v^i, J, F)$. We will omit F and J if it is clear from context what they are. Second component $M_v^i$ of pair $(S_v^i, M_v^i) \in M$ we will call a set **named** $S_v^i$, **$M$** we will call a **named family of sets**.

For a given M and a hierarchical selector U, the map
$\xi \in [\{x1, \ldots, xn\} \to \Omega \cup \Omega^\lambda]$
is called **acceptable**, if for each elementary selection $x \in S_{t(y1, \ldots, yn)}^i$ or $S_x^i \in S_{t(y1, \ldots, yn)}^i$ the relation
$\xi(x) \in M(S_{t(\xi(y1, \ldots, ym))}^i)$
or
$S_{\xi(x)}^i \in M(S_{t(\xi(y1, \ldots, ym))}^i)$
is true.

Other word, the map $\xi$ determines the values for variables that satisfy to the selector U.

For a form F with $S_{t(y1, \ldots, yn)}^i$ in a left side and selector U let's denote $\Sigma_v^F$ all acceptable map $\xi$ that $t(\xi(y_1, \ldots, y_n)) = v$.

**Remark 4.4.1.** The set $\Sigma_v^F$ consists of all data used for a selection data for set $S_v^i$.

The notation $\Sigma_{v|y1, \ldots, yn}$ will be used for a set of values for $y_1, \ldots, y_n$.

**Def. 4.4.2.** The **value for expression** $Q_1x_1 \in S_{t1}...Q_nx_n \in S_{tn}p(x,y,x_1,...,x_n, z_1,...,z_m)$ on $\Sigma_v^F$ defined by induction on number of quantified variables:

**Induction base**: for formula $Q1x1 \in S_{t1}\ p(y_1,...,y_n,x_1)$ value is a value of non quantified formula

$p(y_1,...,y_n, b_1)*...*p(y_1,...,y_n, b_m)$, where '*' is '&' if $Q1 = \forall$ or '*'is '|' if $Q1 = \exists$, $\{b1,...,bn\} = \Sigma_v^F$.

**Induction step**: a formula $Q_1x_1 \in S_{t1}...Q_nx_n \in S_{tn}p(x_1,...,x_n, z_1,...,z_m)$ is equal to $Q_1x_1 \in S_{t1}...Q_nx_n \in S_{tn}p(x_1,...,x_{n-1},b_1, z_1,...,z_m)*...* Q_1x_1 \in S_{t1}...Q_nx_n \in S_{tn}p(x_1,...,x_{n-1},b_m, z_1,...,z_m)$ where * and $b_1,...,b_m$ are defined earlier.

Let's denote as t[M] the set of values of term t on variables values from M.

**Def. 4.4.3.** Family M is called **agreed** with system F on interpretation J, if for each form with left part $S_t^i$ next are true:

a. if F is of type of $\alpha$ and $M_c^i = \{c_1,...,c_k\}$, then $(S_c^i, M_c^i) \in \mathbf{M}$,

b. if F is of type of $\beta$, then $(S_v^i, M_v^i) \in \mathbf{M}$ if and only if $v \in t[\Omega \cup \Omega^\lambda]$, $\Sigma_v^F \neq \emptyset$ and formula of the form is true on $\Sigma_v^F$,

c. if F is of type $\gamma$, then $(S_v^i, M_v^i) \in \mathbf{M}$ if and only if $v \in t[\Omega \cup \Omega^\lambda]$, $\Sigma_v^F \neq \emptyset$, $M_v^i$ is a minimum set with next properties:
   - each element of $M(S_i^{t(d1,...,dn)})$ belongs to $M(S_v^i)$, where $d_1,...,d_n$ are value of parameters of t for $t(d_1,...,d_n)=v$,
   - if for some $d \in S_v^i$ the formula
   $Q_1x_1 \in S_{t1}...Q_nx_n \in S_{tn}p(d0,d,x_1,...,x_n)$
   is true on $\Sigma_v^F$ for some $d_0$, then $d_0 \in M(S_v^i)$,
   - $M_v^i$ does not have other elements.

d. if F is type of $\gamma$, then $(S_v^i, M_v^i) \in \mathbf{M}$ if and only if $v \in t[\Omega \cup \Omega^\lambda]$, $\Sigma_v^F \neq \emptyset$, $M_v^i = [\Omega \cup \Omega^\lambda \cup S(F,J)] \cap t_1[\Sigma_{v|v(t1)}^F]$.

**Def. 4.4.4.** A family **M** is called **selected** by a system F on interpretation J, if M is agreed with F and any family $M_1$ have gotten from M by increasing only one of it sets is not agreed with F. A family of named set also is called **a variant of selection**.

**Example 4.4.1.** Let us have a list of laboratories and its employee's names, years of experience and salaries. These lists may be defined with forms of type $\alpha$:

$Sc=(lab_1,..., lab_k)$,
$S_{lab\ 1}=\{c_1^1,...,c_{n1}^1; lab_1 \in S_c\}$,// employees of $lab_1$
.......
$S_{lab\ k}=\{c_1^k,...,c_{n1}^k; lab_k \in S_c\}$// employees of $lab_k$

Then next form of type $\beta$ defines the set of names of "nice" laboratories where salary is a monotone growing function of experience:

$S=\{\forall S_{lab} \in S \forall x \in S_{lab}((experience(x) > experience(y)) => (salary(x) > salary(y)), S_{lab} \in S_c\}$.

### 4.5. A comparison of subclasses of forms

Basic questions for system of forms are expressive powers of different sub classes. For comparison of expressiveness we will consider systems of definitions as operators from data universe to the family of named sets. The idea is that if for any system of definitions F from a class *A* exists a system with extended signature from another class *B* that generates for any interpretation J the same family of named sets then we may say *B* is not **weaker** than *A*.

We allow extending signature of forms of class B with:
- new variables,
- **secondary predicates** – logical expressions with &¬| operations with symbols of predicates of a form of class A. Because any logical function can be created with these operations, these logical operations may be replaced with any others.

More formal.
**Def. 4.5.1**. Let $\psi^{n1}_1,\ldots, \psi^{nh}_h$ is a function symbols with interpretation $[\{true,false\}^{mi} \rightarrow \{true, false\}]$. Let $p_1,\ldots, p_m$ are predicates symbols. Then **predicate expressions or secondary predicates** we will call such a minimal set of expressions that:
a. base predicates symbols with arguments of terms $t_1,\ldots, t_k$ build above functional symbols and elements of W, are a predicate expressions;
b. if $\psi$ is a secondary function symbol and $p_1,\ldots, p_m$ are predicate expressions, then
$$\psi(p_1,\ldots, p_m)$$
is a predicate expression too.

The example of secondary predicates is p1|((¬ p1(t1)|p2(t2))&|¬ p3(t3)), where p1, p2, p3 are base predicates symbols.

From now and up we will consider systems of form, which predicates could be predicate expressions.

**Def. 4.5.2. Interpretation** J2 is called **similar** to interpretation J1 if $\Omega(J1)=\Omega(J2)$, $\Omega^{\lambda}(J1)=\Omega^{\lambda}(J2)$ and interpretations of their base predicate and functional symbols are the same.
Other word, in similar schemas difference might be in predicate expressions only.

**Def. 4.5.3.** Two systems $F_1$ and $F_2$ with same basic predicates and functions symbols are called **equivalent** regarding to name $S_a$, if for any interpretation J for $F_1$ exists a similar interpretation for $F_2$ that sets with name $S_a$ are the same.

**Def. 4.5.4.** Let $\Phi_1$ and $\Phi_2$ are subclasses of definitions. Then $\mathbf{\Phi_1}$ is called **not less expressive than $\mathbf{\Phi_2}$**, notation is $\Phi_1 \geq \Phi_2$, if for each system $F_1 \in \Phi_1$ and set name $S_a$ exists system $F_2 \in \Phi_2$ equivalent to F1 regarding to $S_a$.
The fact that $\Phi_1 \geq \Phi_2$ but not $\Phi_2 = \Phi_1$, will be denoted as $\Phi_1 > \Phi_2$.

Now we will compare next 6 classes:
$G^c$ is a subclass of systems with formulas containing two variables x and y so that $x \in S_y$ belongs to selector of some form;
$G^{\neg \eta}$ is a subclass of systems without forms of type $\eta$ ($\eta \in \{\alpha, \beta, \gamma, \delta\}$;

G$^s$ is a subclass of systems with a signature extended with predicates of p(x,S) ≡ s∈S.

The main tool for comparison of expression powers of subclasses are special interpretations similar to free interpretations used in [Luckham D.C., Park D.M., Paterson M.S. On formalized Computer programs. //J. Comp. and Syst. Science. 1970, v4].

**Def 4.5.5.** Let F be a system, $\sigma_F^b$ - a set of its functions and predicates. **The standard interpretation** of $\sigma_F^b$ is a pair (Q∪ Q$^\lambda$, D), where

- Q is a finite set of elements of $q_i$, i∈N,  called **bearer**;
- Q$^\lambda$ is a finite set of terms build of functional symbols of $\sigma_F^b$ with arguments of Q;
- a diagram D is a finite set of strings "$p_k(t_1,…, t_k)$" or " ¬$p_k(t_1,…, t_k)$" where p is a predicate symbol, {$t_1,…, t_k$} ⊂ Q∪Q$^\lambda$, and if "$p_k(t_1,…, t_k)$" is in D then "¬$p_k(t_1,…, t_k)$" is not in D and inverse, if "¬$p_k(t_1,…, t_k)$" is in D then "$p_k(t_1,…, t_k)$" is not in D.

- the value of function $f^{\,k}∈ \sigma_F^b$ on {$a_1,…, a_k$} ⊂ Q∪Q$^\lambda$ is an string "$f(a_1,…, a_k)$" .

- the value of predicate $p^k∈ \sigma_F^b$ is next:

$$p(t_1,…,t_k)= \begin{cases} \text{true, if the string ``}p(t_1,…,t_k)\text{''  is in D,} \\ \text{false, if the string ``}\neg p(t_1,…,t_k)\text{''  is in D,} \\ \text{void, in other cases.} \end{cases}$$

**Example 4.5.** 1. Let Q={$q_1$, $q_2$, $q_3$}, D={p($q_1$,$q_1$), p($q_1$,$q_2$), p($q_2$, $q_2$), p($q_3$,$q_2$),p($q_3$, $q_3$)}. Then a system $F_0$ with the only form {S={∀x∈S∀y∈S p(x, y)} defines two variants of sets: M(S)={$q_1$},
M(S)={$q_2$,$q_3$}.
For the first case [∀x∈S∀y∈S p(x, y)] = p($q_1$,$q_1$), for the second case
[∀x∈S∀y∈S p(x, y)] = p($q_2$,$q_2$)&p($q_3$,$q_3$)&p($q_2$,$q_3$)&p($q_3$,$q_2$).

Both are true on the standard interpretation.

**Def. 4.5.6.** For an element $q∈$ Q∪Q$^\lambda$ **the collection of its properties** is called a set of expressions of D containing $q$. This set we will denote as φ($q$).
For previous example φ($q_1$)={p($q_1$, $q_1$), p($q_1$,$q_2$)}.

**Def. 4.5.7. Full collection of properties** of element q∈ Q∪Q$^\lambda$ is defined as a result of next process:

- Each element $q_p$ from arguments of φ(q) are replaced with the string $q_p$[φ($q_p$)] if $q_p$≠q;
- Repeat the rule above for each new elements appearing on previous steps.

For an example above, the full collection of properties for $q_1$ is
$q_1$[p($q_1$,$q_1$), p($q_1$, $q_2$[p($q_2$, $q_2$), p($q_2$,$q_3$[p($q_3$,$q_3$), p($q_3$,$q_2$)])])].

**Def. 4.5.8.** The standard interpretation J of a system F we will call **minimal for set $S_a$** if for each such standard interpretation $J_1$ that $Q(J_1)=Q(J)$, $Q^\lambda(J_1)=Q^\lambda(J)$, $D(J_1) \subset D(J)$ and $D(J_1) \neq D(J)$, take place $M(S_a,J_1,F) \neq M(S_a,J,F)$.

Other word, a removing some elements from diagram of a minimal standard interpretation changes the set with name $S_a$.

**Lemma 4.5.1.** If for a system $F_1$ minimal for $S_a$ standard interpretation is not minimal for the same set $S_a$ for a system $F_2$, then $F_1$ is not equal to F2 for the set $S_a$.

**Proof.** Let t be an element of D(J) and removing it does not change the $M(Sa,J,F_2)=M(Sa,J_1,F_2)$. Let $J_1$ is made from J by removing t from D(J), then $M(Sa,J,F_2)=M(Sa,J_1,F_2)$ and $M(Sa,J,F_1) \neq M(Sa,J_1,F_1)$, because J is a minimal for $F_1$. If we suppose that $F_2$ is equal to $F_1$ on Sa, then $M(Sa,J_1,F_1)=M(Sa,J_1,F_2) =M(Sa,J,F_2)= M(Sa,J,F_1)$, that contradicts to $M(Sa,J,F_1) \neq M(Sa,J_1,F_1)$. End of proof.

**Def. 4.5.9.** Let J be a standard interpretation of F, $\psi_1,\dots,\psi_k$ be a logical functions. **The secondary for standard interpretation** J is called an interpretation $J^{sec}(J)= (Q \cup Q^\lambda, D^{sec})$, where $D^{sec}$ is the extension of D with expressions "$\psi_i(t_1,\dots t_p)$" if it is true on D or with the expression "$\neg \psi_i(t_1,\dots t_p)$" if it is false on D. D' is called a **secondary diagram**.

Let's **$\min_{Sa} J^{sec}(J)$** denote a minimal for $S_a$ interpretation $(Q \cup Q^\lambda, D^{min})$ for which $D^{min} \subset D^{sec}$.

**Remark 4.5.1.** For each element q replacing logical functions with a disjunctive normal formula with only basic predicates we will got expression as
$(r_1^1(t_1^1)\&\dots\&r_i^m(t_i^m))|\dots|(r_j^1(t_j^1)\&\dots\&r_j^m(t_j^m))$,
where r is p or $\neg$p. This expression is true if at least for one s $q[r_s^1(t_s^1)\&\dots\&r_s^m(t_s^m))]$ is in the diagram D.

All next results of comparison classes are based on a structure of collections of properties of bearer's elements.

One of the basic results of the theory of models in mathematical logic is that names and secondary predicates can be eliminated. So there they are used only for short. Next result shows that if to use a comparison defined in **Def. 4.4.3,** then a hierarchy of names increases the expressive power of system of forms.

**Theorem 4.5.1.** Let $G^n$ to be a class of forms with no parameters in set names, $G^{-n}$ to be a complimentary $(G^{-n} = U-G^n)$ set to $G^n$. Then $G^{-n} < G^n$.

**Proof**. The truth of $G^{-n} \leq G^n$ is come from the fact that by adding a form with a new type of parameter names to any system from $G^{-\alpha}$ creates a system from $G^\alpha$ equivalent to original.

Now we will prove that $G^{-n} < G^n$ strictly.

Let's $F_0$ be a system of two forms

$S_0=\{\forall S_x \in S_0 \; \forall y \in S_0 \; \forall z \in S_x \; \forall u \in S_y \; r(z,u)\}$

$S_z=\{\forall x \in S_z \; \forall y \in S_z \; p(x, y, z); z \in S_1\}$

One can see that collection of properties of an element q of a set $S_z$ on a minimum interpretation has a form

$q[\varphi: q \in S_z \in S_0] = q[\varphi: q \in S_z, r(q, q_1^{-1}[\varphi: q_1^{-1} \in S_{z1} \in S_0], \ldots, r(q, q_1^{-1} \in S_{zn1} \in S_0]), \ldots, r(q, q_1^m[\varphi: q_1^m \in S_{zm} \in S_0], \ldots r(q, q^m_{nm} \in S_{zm} \in S_0))])]$, where $\{q_1^j, \ldots, q^j_{nj}\} = M(S_z^j)$ for $S_{zj} \in S_0$; m is a number of names $S_{zj}$ from $S_0$.

Here the record $q[\varphi: q \in S \ldots]$ is for short of the list of properties that determine that $q \in M(S)$.

Replacements them with full records gives a record containing $< h^h$ base predicate r, where $h = \Sigma n_{zj}$.

For a system from $G^{\neg n}$ sizes of collections of elementary properties are less than production of sizes of sets under quantifiers of any formula. Size of collections for elements of these sets also restricts the production of sizes of other forms.

Let's add to the bearer Q new elements and to D such properties that increases each $M(S_{zi})$ $(S_{zi} \in S_0)$ independently for $d_{zi} \gg n_{zi}$ elements. Then a size of properties in collection for q will be proportional to $h^h$, where $h = \Sigma d_{zi}$.

For any of systems L from $G^{\neg n}$ the size of properties could not increase more than $\Sigma d_{zi}$. Number of sets used for forms $F_j$ in L is a constant $k_j$, number of form is a constant so number of properties in property collections for any element is not exceed $C_0 * d_{zi}^{C1}$ for some constants $C_0$ and $C_1$. So selecting $h > C_1$ we may conclude that in $G^{\neg n}$ no system with such properties. Proof is finished.

With the same technique can be proved the

**Theorem 4.5.2.** Let $G^r$ is a subclass of systems where each set name $S_a$ has a constant low index $a$ and hence number of all set names is a constant. Then $G^r < G^c$.

Let G be a class of all systems. Then next statement is true.

**Theorem 4.5.3.** $G^{\neg \beta} < G$.

Proof. We will proof that for system with one $\beta$-form
$$S_0 = \{\forall x \in S_0 p(x)\}$$
there is no equivalent in $G^{\neg \beta}$.

Suppose that $S_0$ can be be defined in $G^{\neg \beta}$. Then this form cannot be $\alpha$ type, because sets of properties for its elements for minimum interpretation are empty. It cannot be $\delta$ type because $S_0$ does not have any functions in it definition.

Last case is if $S_0$ defined with form $\gamma$.

Let $S_b$ be a base set of the definition. If $S_b$ is defined with form of type $\alpha$ then $S_b$ has an elements with empty set of properties and hence assumption that $S_0$ can be defined in class $G^{\neg \beta}$ wrong. If $S_b$ is defined by form $\gamma$ then consider its base set and so on while all forms from F will be exhausted. Proof is finished.

Similar statement is true for $\gamma$ type of definitions.

**Theorem 4.5.4.** $G^{\neg \gamma} < G$.

**Proof**. Let F is a system with two forms
$S_0 = \{c_1, \ldots, c_k\}$
$S_g = \{S_0, p(x,y), x \in S_g, y \in S_g\}$.

So $F \in G$ and $\neg(F \in G^{\neg \gamma})$.

Suppose the theorem is not true, there is a system F1$\in$ G$^{\neg\gamma}$ and F1 define the same set S as F. Let's consider interpretation J with a special diagram D constructed with next way. Let q[…p($q_j$,$c_i$)] is an extended representation of element of M(J,F,S), its deepest element has to be from S0.

The deepest element (let's denote it $q_b$) in F1 has to be from S, be an argument of p($q_a$, $q_b$)  and be under existential quantification because in other case there is no deepness bigger than summation of sizes of all forms. Deepness for F does not have restriction. Then if q[…p($q_j$,$c_i$)] is an element of set for S in F, then replacing p($q_j$,$c_i$) with  true p($q_j$,q) – because "p($q_j$,q)" belongs to a diagram - makes element q[…p($q_j$,q)] to belong to S.

Let's add to the diagram D a property p($q_j$,q) for each p($q_j$, $c_i$).

Then q will have a loop without elements of $S_0$ and a set of elements from this full collection of properties  is a selected set for F1 but not for F. Now if we add copy of elements M(J,F,S)-{$c_i$} and properties copied from q[…p($q_j$,q)] with replacing elements to its copy we will get new interpretation $J_w$ where F1 has two sets for S, but F has only one. Hence on interpretation $J_w$ systems F1 and F have a different variants of the selected sets for S. End of proof.

**Remark 4.5.1.**  The proved theorem states that an induction can't be expressed with a recursion. The way of a proof also shows that a reason is in variants: an induction based on minimum sets, a recursion based on maximum sets.

The request that the selected set has to be maximized has downside discribed in the next statement.

**Theorem 4.5.5.** Class G$^{\neg\gamma,\delta}$  has systems with many **agreed** sets but named family with maximum set for some name does not exists.

Proof.  Let's consider system with two forms:

$S_1$={($\forall x_1 \in S_1 p(x_1)$)&($\forall y \in S_2 \exists z \in S_1 r_1(z,y)$)}

$S_2$={($\forall x_2 \in S_2 p(x_2)$)&($\forall y \in S_1 \exists z \in S_2 r_2(z,y)$)}.

Let for some interpretation M1 and M2 to be **agreed** subsets. The maximum set for S1 is $M_1{}'$={x|p1(x)}, but this set can be satisfied with second only if formula

$\forall y \in M_1{}' \exists z \in S_2 r_2(z,y)$

is true with this $M_1{}'$. Clearly, that exist interpretations where this formula is not true.

**Theorem 4.5.6.** Let's enriched signature of systems with interpreted predicate x$\in$S and denote its class G$^s$. Original class of systems let's denote G. Then

G<G$^s$.

**Proof.** In systems from G it is impossible to define a complementary set for the set defined with numerated form. In G$^s$ it is obviously possible.

**Discussion.**

The results shown here are just the beginning of studying of nonprocedural definition of subsets. Possible questions for further researches might be new forms of definitions and a **set naming**. So far we assume that names are given independently of the sets elements, but if to make sets names dependent of its elements does it increase the expression power?

**4.6. The complexity of algorithms of building defined sets**

Results of the previous paragraph show that systems of definitions of named set is a powerful tool for a program specification. The main obstruction to use it as a programming language is a still open satisfiability problem.

But for a wide enough subclasses fast, at least polynomial algorithms exist.

We will see that some additional info about bearers and predicates helps to reduce complexity to polynomial or even leaner dependency.

For instance if for definition

S={∀x∈S∃y∈S p(x,y)}

it is known that p is asymmetric predicate (p(x,y)= ¬p(y,x)) , then M(J,F,S) =∅ for any F and J.

Another simplification comes from knowledge that for each element x of S exists only one element y of S so that p(x,y) =true.

The leaner algorithm is next.

Let's build a sublist of pairs

P={(x,y)|p(x,y)=true}

with the next way. Take a first element of a given finite start data (universe), let it be **a**, and make next step. Find first not selected element **b** that p(**a**,**b**)=true. If there is no such element, then add pair (**a**, ∅) to P, and **a** is called dead-end element. Otherwise repeat the step. When all elements will be checked, the set of non dead-end elements from P is a set for S.

**Remark 4.6.0.** Let's note an important feature of selected set. It was defined as a set which **can't be increased by adding only one element**. The next example shows that it may exist two variants M1 and M2 so that M1 ⊂ M2, but M2 can't be constructed by adding to M1 elements one by one.

The example system contains one form

S={∀x∈S∀y∈S p(x,y)},

universe is

V={a,b,c,d,e,f},

diagram is

D={p(a,b), p(b,a), p(c,d), p(d,c)}.

Then M1={a,b} and M2={a,b,c,d}.

Clear M3=M1∪{c} or M4=M1∪{d} are not a variant for S.

Let's divide systems of form for subclasses with different number I of form types, number J of quantifications{∀, ∃}, number of their changes K and identify them with vector (I,J,K). So subclass (1,1)=(1,1,0) includes systems with one type of forms and one quantification (0 quantification changes). Algorithms for selection subsets for them are straightforward.

The subclass (1,2,0) consists of forms of next 4 types:

1) S={∀x∈S∀y∈S p(x,y)}
2) S={∀x∈S∃y∈S p(x,y)}
3) S={∃x∈S∀y∈S p(x,y)}
4) S={∃x∈S∃y∈S p(x,y)}.

Farther we will consider a complexity related to the size of data universe.

**Lemma 4.6.1.** For subclass (1,2,0) exists a polynomial algorithm of data selection.

**Proof.** For a system of type 1) if a∈M(S) then p(a,a)=true and hence at least one element set can be found for linear time or detect of it absents. If universe V contains such element v, that for already constructed part M for each x∈M

p(x,v)&p(v,x)&p(x,v)&p(v,v)=true,

then v may be added to M. The part M, which can't be extended such a way, is a variant for S.

**Remark 4.6.1.** The algorithm can be generalized for any number r of existential quantifications. Here a selection of the first element **a** for M is made with condition p(**a,…,a**)=true and at each step k+1 set $M_k$ is extended with element v if

∀$y_1$∈$M_k$∀$y_2$∈$M_k$…∀$y_r$∈$M_k$(p(v, $y_1$,…, $y_r$)&p($y_1$,v,$y_2$,…,$y_r$)&…&p($y_1$,…,$y_r$,v)).

For a system of type 2) the algorithm creates a subset of universe by removing on each stages elements that is for sure not in M(S). On stage 1 algorithm removes such x∈V that p(x,y)=false for any y∈V. Notes that if V does not have such elements then whole V is a M(S). On second and on any next step k algorithm remove elements with the same properties but with universe

V-{$X_1$∪…∪$X_{k-1}$},

where $X_i$ are elements removed on stage i.

The process of removing is finished when all elements are removed or no one was removed on current stage. Number of stages does not exceed the size of universe, because at least one element has to be removed.

Proof of algorithm.

Let's call element y for which p(x,y)=true, a **supporter** of element x. Then on stage 1 algorithm removes elements that do not have supporter in any subset of universe. If there are no such elements – any element has a supporter - then a whole universe is a set M(S). On next stages algorithm remove elements which has only supporters that do not have supporters for themselves.

Let's proof that the rest is M(S).

Let's suppose there is an element v without support in the rest, meaning its support was removed. But by construction v also have to be removed. The contradiction finishes the proof of current case.

**Remark 4.6.2.** The algorithm may be extended for the form S={∀x∈S∃y∈S…∃z∈S p(x,y,.., z)} defining supporters for x a set of several elements y,…,z.

**Remark 4.6.3.** Similarly can be build a polynomial algorithm for more general cases of forms with one universal quantification and existential others.

**Remark 4.6.4. Classes 2) and 4) have a remarkable properties:**
- they have only one variant
- the complexity of the algorithm above is $n^2$,
- such type of definitions are close to the induction.

**Remark 4.6.5.** Definitions on the type 2) and 4) can be written as a system of Horn's disjunctions.

Sets for system of type 3) can be built with extensions of the set for form
$S_a$={∀y∈$S_a$(p(a, y)&p(a, a))}

with a next way. In a current set $Ma =M(Sa)$ find such element b that $p(b, x)$ is true for each $x \in Ma$ and $(V-Ma)$ has an element y that $p(b, y)$ is also true. The set $Ma$ is extended with this element y and extension steps are repeated again.

Systems of type 4) give empty sets if $p(x, y)$ is false for any pair $(x, y) \in V \times V$ or the whole set V if $p(x, y)$=true for some pair from V.

End of proof of lemma 4.6.1.

For class (1.3.0) we already considered cases 1-4, so we have to consider systems

5) $S=\{\forall x \in S \forall y \in S \exists z \in S\ p(x, y, z)\}$

6) $S=\{\forall x \in S \exists y \in S \forall z \in S\ p(x, y, z)\}$.

**Lemma 4.6.2.** For interpretations of systems from classes (1.3.1) and (1.3.2) where for each variable under $\exists$ exists not more than one element to satisfy $p(…)$, exists a polynomial selection algorithm.

**Proof.** Non trivial systems from classes above are

$S=\{\forall x \in S \forall y \in S \exists z \in S\ p(x, y, z)\}$

and

$S=\{\forall x \in S \exists y \in S \forall z \in S\ p(x, y, z)\}$.

For the first one let's consider interpretations satisfying to lemma: if $p(x,y,z_1)$=true and $p(x,y,z_2)$=true then $z_1=z_2$. Note that if $a \in S$ and if $p(a,a,b)$=true for an element b then $b \in S$, because such element is unique, no another element could support a. So any element b that $p(x,y,b)$=true and x,y are in $M(S)$ also has to be in $M(S)$.

This gives the idea of the algorithm for building selected set.

Take elements **a** and **b** that $p(\mathbf{a,a,b})$=true  and add **b** to current set $Ma$. Then add to $Ma$ any element **c** that $p(b,a,c)$ or $p(a,b,c)$ or $(p(b,b,c)$. If all such elements are already in $Ma$ - stop the building. If there is no such **c** − then $Ma$ is empty and algorithms starts with new **a**. If all **a** were tried then stop − selected set is empty.

The final set $Ma$ obviously satisfies to the form and a complexity of this algorithm is $O(n^3)$.

For a system of second type one of sufficient conditions for existing of a polynomial algorithm is the non emptiness of variants for more strictly form

$S1=\{\forall x \in S1 \exists y \in S1 \forall z \in V\ p(x,y,z)\}$ (V is an universe).

Obviously that if $M(S1)$ is not empty, then it satisfy for original form

$S=\{\forall x \in S\ \exists y \in S\ \forall z \in S\ p(x,y,z)\}$.

If to use a predicate $p1(x,y)=\forall z\ p(x,y,z)$ then form

$S1=\{\forall x \in S\ \exists y \in S\ p1(x,y)\}$

defines the same variants and belongs to class (1.2.0) which is polynomial.
End of proof.

Another **sufficient condition** is a commutativeness of third and first or second arguments of a predicate p. In this case applicable algorithm for system (1,2, 0) with universal quantifications $\forall$ only.

**Third sufficient condition** is if an argument under existential quantification is a (Scolem's)  function of one or all others arguments. In this case we again may replace 3-D predicate with 2-D and formula has only $\forall$ quantifications.

**Forth sufficient condition** is if a predicate has separateable variables:

$p(x,y,z)=r(x,y)*t(z)$.

For the case of '*'='&' let's select such elements z of universe V that $t(z)$ = true. Denote this set of elements as V1. The form 5) is equivalent on V1 to some form of class (1,2) :

$S_1=\{\forall x \in S1 \ \exists y \in S1 \ r(x,y)\}$

Its building algorithm is polynomial as it was shown earlier.

For the case of '*'='|' a selected set is a union of sets for

$S_1=\{\forall z \in S_1 t(z)\}$

and

$S_2=\{\forall x \in S_2 \exists y \in S_2 \ r(x,y)\}.$

**Fifth sufficient** condition is if universe V may be factorized for small number of sets $V_1,\ldots,V_r$ (say r independent or logarithmically dependent of the universe size) for which

$\forall u_1 \in Vj \ \forall u_2 \in Vi \ \forall y \in V \forall z \in V \ (p(u1,y,z) \equiv p(u2,y,z)).$

In this case

$\forall x \in S \ p(x, y, z) \equiv p(a_1,y,z) \& \ldots \& p(a_r, y,z)$

where $a_1 \in S \cap V1, \ldots, a_r \in S \cap V_r$ are arbitrary elements. Because r is a constant form 5) is equivalent to form (1,2,0).

Another way to build selected sets is to use simpler form to define **approximation of sets**. For the class 5) form

$S=\{\forall x \in S \ \exists y \in S \ \forall z \in S \ p(x,y,z)\}$

Such approximation the form

$S_1=\{\forall x \in S_1 \exists y \in S_1 \ p(x,y,x) \& p(x,y,y)\}.$

Clear, that M(S)=M(S1). If M(S1) is small then set M(S) may be found by direct checking. Also we may work with smaller universe V=M(S1), where

$S=\{\forall x \in S \ \exists y \in S \ \forall z \in S \ p(x,y,z); \ x \in M(S1), y \in M(S1), z \in M(S1)\}.$

### 4.7. Fuzzy sets

A natural extension of a set definitions is using fuzzy values from interval [0,1] instead of two-element value {0,1} for belong relations. This extension is used for data mining, objects taxonomy, determine semantics of a text parts.

There are several ways to combine this notion with set definitions.

First is to use fuzzy predicates in set definitions, second – to use fussy quantification, third - to use fuzzy function for predicate of "x is an element of S" used in forms, forth – to use number of elements of sets with some relations as a measure of belonging to the set, fifth and so on may be combinations of these ways.

**Example 4.7.1.** Fuzzy predicates are color of the objects – pale red, dark red, or text relation to the subjects – main issue, just mentioned, ….

**Example 4.7.2.** Fuzzy quantifications $W^t x \in S \ p(x)$ may be used for an expression like "more than t=10% of elements of S have to have a property p(x)".

**Example 4.7.3.** Fuzzy relation $x \in S$ for hierarchical selectors in forms may be useful for selecting elements with a strong relation to some set.

**Example 4.7.4.** Fuzzy measure of a weight of element x in the set by counting of elements y of this set having relation p(x,y).

These are new and rich objects for researches. To select the most fruitful direction probably better to start with applications like unstructured text analysis.

## 4.8. Systems of logical equations for named sets

Let's for each data **d** from universe and each set name $S_i$ map a logical variable $ld_i$. It is true if $d \in S_i$ and false if not.

Then form of definition for $S_i$ generate a system of logical equations

$ld_1 \& \ldots \& ld_n = true$.

The solutions of the system are elements for selected sets.

The complexity of this task is not less than complexity of satisfiability problem because the task contains them. But some heuristic algorithm for solving this problem for O(n2) time for 90% of randomly generated systems exists.

## Chapter 5. The non-procedural language for business data processing

### 5.1. Introduction

A data processing can be seen as a step by step simultaneously transformation elements of some named sets of data.

For scientific needs, as we saw in the chapter 2, these sets are linear hyperspaces in multi-dimensional arrays. Business data have more complex structures and are needed another type of descriptions.

In this chapter we will use systems of data set definitions not only for selections but for processing them as well. For this we will use **open systems** of definition when some of sets considered as 'a priory' given sets, not defined with a current system of definitions. Programs are finite sets of open definition systems. One system is declared as a start one. On each next step are applied systems with updated on previous step some of open sets.

So we are using data control instead of direct control as in most of modern languages. Data driven processing have its pro and cons. Main advantage is that adding new functionality is simple – just add one more open system of definitions. It can be done without recompiling or even postponing current run. Debug also local – if input-output is the same then result also did not change. Disadvantage is less usage so far in practice.

### 5.2. Formal systems for non-procedure data processing

From now up we will consider open systems of data sets definitions. Type of names of input sets for system F are denoted I(F), others types of names are denoted O(F). Open system F is an operator from one family M of named sets with names from I(F) to another family of sets G with names from O(F), notation G=F(M).

G may contain new names and elements if F has of forms of type $\delta$.

**Def. 5.2.1. Data processing specification** (**DPS**) is a pair (**F**, **S**) where **F** is a finite set of systems, **S** is a set of names for (start) data sets. DPS goes by steps. For any step k>=1 DPS execution builds a family of named sets $F_i(G_{k-1})$ for those $F_i$ whose input is in start data sets $G_0$ and is in $\Delta = \bigcup_{0<i<p} F_{jp}(G_{k-1})$ if k>0. The process is finished when for each system some of input sets are empty or no one was updated on previous steps.

Generally the process is **non determinative**; result depends of an applying order of initialized systems.

**Example 5.2.1.** (summation). Let's use some interpreted functions:
- **Abit**(S,k) creates non intersected k-element sets arbitrarily selected from S,
- **plus** (Sa) creates x+y for two elements set {x, y}.

Then the DPS G = {$F_1$:   $S_1$={ **plus**($S_a$); $S_a \in$**Abit**($S_1$, 2) }} calculates sum of data of set $S_1$. $F_1$ divides set $S_1$ to pairs, calculate sum for each of them and return them to set $S_1$. At next steps it again divides $S_1$ to pairs and so on. The process is finished when S1 has less than two elements.

Next two paragraphs shows functional fullness and control power of DPS.

**5.3. Functional fullness of DPS**

Here we will prove that any partial-recursion function (pr-function for short) in Kleene's formalization can be calculated with some DPS.

For variable $x_i$ in pr - formalization corresponds one-element set with a value of $x_i$. Notation f[$S_{x1} \times ... \times S_{xn}$] means that f is applied to the Cartesian product of elements of  $S_{x1}, ..., S_{xn}$ and it equal f($x_1, ..., x_n$).

Then DPS $O_f$[$S_{x1}, ..., S_{xn}$]$\rightarrow S_z$ for V=N,$V^\lambda$=N we will call **representation of function** f($x_1, ..., x_n$)[$N^n \rightarrow N$], for start set Sx1={<$x_1$>}, DPS stops if and only if f($x_1, ..., x_n$) not void and $S_{x1}$ contains a value m=f($x_1, ..., x_n$).

**Theorem 5.3.1.** A class of DPS with V=N, $V^\lambda$=N, where N is a set of natural numbers, with interpreted functions o()$\equiv$0, s(x)$\equiv$x+1, $t^n_m$($x_1, ..., x_n$)$\equiv x_m$ and predicates  '=' and '<' contains representations for any pr-function.

**Proof.** We will base on Kleene's formalization of pr-functions with basic functions o(), s(), t() and operators of compositions**,** primitive recursions and minimization. Proof will be done by induction of number k of operators in pr-function.

For k=0, f is one of basic functions and it representations are:
$S_z$={o[$S_x$]},             //return 0 for any argument,
$S_z$={s[$S_x$]},             //return x+1
$S_z$={$t^n_m$[$S_{x1}, ..., S_{xn}$]}  //return m-th argument.

Let's suppose statement is true for any k$\leq$n and proof that it also true for n+1.

If f constructed with n+1 operators, then there are three possibilities:
1)  f is defined with the last operator of composition
$$f(x_1, .., x_n) = g(h_1(...), ... h_m()),$$

2)  f is defined with the last operator of primitive recursion

$$|f(0,x_1, ..., x_n) = g(x_2, ..., x_n), \text{ if } y=0,$$
$$|f(y+1,x_1, ..., x_n) = h(y+1,x_1, ..., x_n), \text{ if } y>0,$$

3)  f is defined with the last operator of minimization

$$f(x_1,\ldots,x_n)=\mu(y,g(x_1,\ldots,x_n)).$$

Functions g and h contain less operators than f, hence not more than n operators and by induction assamption they have a representations $O_g$ and $O_h$ in DPS class. Then f in case of composition can be expressed as DPS

$$S_z=Oh(O_{g1}[S_{x1},\ldots,S_{xp}],\ \ldots O_{gp}[S_{x1},\ldots,S_{xp}]),$$

in case of  primitive recursion - as two systems $F_1$ and $F_2$:

$F_1$: $\{S_z=Og[S_{x1},\ldots,S_{xp}],\ S_i=\{0\}\}$
$F_2$: $\{S_i=\{s([S_i]);\ \forall v\in S_y\ \forall u\in S_i(u<v))\},\ \ S_z=Oh(S_i,\ S_z,\ S_{x1},\ldots,S_{xn})\ \}.$

Really, the system $F_1$ is applied only once at the beginning, because set it uses will never changed. $F_2$ can be applied only after $F_1$ because at the beginning sets $S_i$ and $S_z$ are empty. But it will be applied on each next steps because $S_i$ will be overwritten and contain number 0,1,2,..., y. Set $S_z$ will get values $f(0,x_1,\ldots,x_n)$, $f(1,x_1,\ldots,x_n),\ldots,$ $f(y,x_1,\ldots,x_n)$.

When a value of element $S_i$ reaches a value y, on next step it will be $\emptyset$ and $F_2$ can't be applied.

In the case of minimization, DPS consists of two systems $F_1$ and $F_2$:
$F_1=\{\ S_i=\{0\}\},$
$F_2=\{\ S_i=\{s[S_i];\ \forall y\in S_u(y\neq 0))\},\ \ S_u=Og(S_i,S_{x1},\ldots,S_{xn}),\ S_z=o[S_i]\}.$
$F_1$ will be applied only once at the beginning.

$F_2$ will be applied on each next steps while $S_i\neq\emptyset$. It stops when $g(i,x_1,\ldots,x_n)$ is not defined and hence $S_u=\emptyset$, or when $y=0 \Leftrightarrow g(i,x_1,\ldots,x_n)=0$ and i is an element of the result set $S_z$.

End of proof .

### 5.4. Data control and Petri net

Petri net is one of best distributed system model combining power, simplicity and clear visual presentation. The target of the paragraph is to show that data control used in DPS is not weaker than Petri nets.

**Petri net** is a bipartite directed multi-graph with vertexes set $P\cup T$ ($P\cap T=\emptyset$), P is a set of **positions**, T is a set of **transitions**.  **Marking** is a function $\mu:[P\rightarrow N]$, N – natural numbers. A transition t is called **allowed** if for each incoming position p $\mu(p)\geq k(p,t)$, where $k(p, t)$ is a number of edges between t and p.  **An initialization of transaction** t with marking $\mu$ is called a replacing  $\mu$ with $\mu'$: $\mu'(p)=\mu(p)-k(p,t)$ for input p and $\mu'(q)=\mu(q)+k(t,q)$ for output q of transaction t.

By introducing for each vertex a set name and constructing for each transaction t a system of definitions it can be modeled Petri net even with **resistant** edges.  But inverse is wrong, not each DPS can be modeled with Petri set. The proof is very technical but strait forward. So we omitted it.

### 5.5. Practical language based on systems of definitions

### 5.5.1. Introduction

The target of this section is to evaluate the practical usefulness of system of definitions. The next system called Dictionary Driven Reports (DDR) with a business data processing language shows how the system of named sets of data can be used for application domain description, how much semantics it may carried even if its predicates are simple and how small additional info are needed to describe data fullness and to specify algorithms to process.

Such systems might be useful for data mining and an unstructured data processing. A system of tables is what holds main and easy to process information in economic, financial, scientific and engineering texts. The value of tables cannot be evaluated without knowledge of a system behind them.

**The first idea** considering here is that systems of tables are defined via four hierarchical named lists: list of names for the particular type of tables, hierarchical list of columns names, hierarchical list of names for rows and list of table attributes which carry context data related to the all data in table.

Type and name fully identify the table instance. One type table with multiple names we also call a **table series**.

**The second idea** is that a table processing is not an arbitrary – output tables have to be homomorphism images of input tables. If there is no "natural" homomorphism – then one table can't be got from others. Other word, the representation determines semantic of the tables.

**The third idea** is that cells of input tables that mapping to one output cell might be processed independently in parallel.

So it is an example of parallel programming without using a word "parallel" or "synchronization".

For the next table reflecting office expenses we have seven lists

DEPARTMENT={IT, HR},
QUARTERS={1,2,3,4},
YEARS={2007,2008},
PROJECTS={pr1,pr2},
EXPENSE={office, trip}.
PERSONEL={p10,p11,p21,p22, p23}
INDICATOR={expenses, personnel, amount, max personnel, max amount}.
ADDRESS = {zip code}.

Then expenses table is:

names=DEPARTMENT$\times$QUARTER$\times$YEAR
columns=INDICATOR
rows=PROJECTS//EXPENSE//PERSONEL
table attributes = ADDRESS$\times$ DEPARTAMENT $\times$ QUARTER $\times$ YEAR.

Next is an example of such table:

type: EXPENSE REPORT
name:-
table attribute:

| Address | Depart | quarter | year |
|---------|--------|---------|------|
| 08550 | IT | 3 | 2007 |

Row/columns:

| projects | amount |
|----------|--------|
| pr1 | $12 |
| office | $8 |
| p11 | $3 |
| p01 | $5 |
| trip | $2 |
| p11 | $2 |
| pr2 | $7 |
| office | $4 |
| p21 | $4 |
| trip | $3 |
| p23 | $3 |

For a table definition it can be used expressions with regular list operations and some additional ones. In the example above table attributes defined as a Cartesian product of four lists ADDRESS×DEPARTMENT×QUARTER×YEAR, list of rows is constructed with a special operation '//' to get an ordinary list from hieratical lists PROJECTS//EXPENSES//PERSONAL.

If we have a table and its lists description we may check if personal IDs are correct, departments and date are OK. We may check if we got all reports from each department and for current quarter and year. Also we may check if its processing is correct using additional info about **Suppes - Zinnes measurement scales** – set of operations which make sense to do.

For processing column "amount" it make sense summation, multiplication to constant (to convert one currency to another), but not allowed multiply amount $A to amount $B.

It make sense a square feet but does not make sense a square dollars or kilograms. But the operation of division amount to amount is made sense.

Here the DEPARTMENT list has a simplest scale – individual, for which only operations $=, \neq$ and counting elements are allowed. The QUARTER list has a comparison scale: in addition to previous operation it has a $<, >$ operations. The column "amount" has a scale with all operations above and also multiplication to constants, dividing to number from personnel column, summation, max, min and so on.

For a hierarchical list of rows we allow summation operations only for all items inside one level hierarchy, but do not allow mix different levels or use part of them. Such sort of information has a great deal of semantic and helps to build **intellectual user interface**.

The transposition of columns and rows does not change the information in tables and processing algorithms.

For **data mining** algorithms it gives targets for further search: how tables are constructed, what lists have to be found, what is the (operational) semantics of data, how to check a completeness of data (if tables or some part of table are missing) and their correctness.

For a **data verification and data completeness** the description allow to check if all reports from each department and current quarter are received (list of departments DEPARTMENT and list of quarters QUARTER) and to avoid misspelling user has only to select. The list of absent reports can be generated as well. Also a report for the same quarter of last year can be defined. For additional data checking it can be used a test that numbers are not differ for more than 15% from numbers of the previous report.

The important feature of the LDR is that data has a human representation – two dimensional tables with 4000 years of experience and polishing.

Further we will use more specific type of lists and call them **dictionaries**.

### 5.5.2. Dictionary Driven Reports (DDR) application for business data processing

Core of the DDR system are notions of local universes (sets of elements for dictionaries and tables), dictionaries (special type of named lists with elements from some universe) and tables. Examples of local universes are a list of employee names, list of cities in USA. Dictionaries are subsets of these universes. Example of such subset is a list of IT department employee's names. For minimizing input and avoiding misspelling subsets are identified with universe's name and set of indexes of its elements in this universe.

There are "a priory" given local universes

- NaturN is a set of nutural numbers,

- RealR is a set of real numbers,

- CharC is a set of strings of characters

- Intervals is set of pairs of numbers or strings.

More complex types of universes are vectors and graphs.

Table definition lists might be defined as results of list operations.

**Def. 5.5.2.0**. A **dictionary** has a **name** (string of characters) and consists of two components:

- name of local universe,

- list of indexes for elements of this universe accompanied with genesis - sets of dictionary names (and hence a universe name) to which the element belongs.

DDR applications have a set of operations operS - a 3D vector (function name, list of local universes for arguments, list of local universes for results).

The operation is **applicable** if input cell dictionaries contain dictionaries in operation arguments and output cell contains output dictionaries.

DDR Design Studio consists of several subsystems. One of them is a dictionaries and tables definitions subsystem. It supports inputting named dictionaries, creating new dictionaries with operations and defining tables with four dictionaries – dictionary of names, dictionary of table attributes, dictionary of column names and dictionary of row names. The subsystem checks if all of them are filled up and if not – force to do it. If all dictionaries are filled up, it checks if all tables with possible names are filled. If not – gives a list of absent tables.

Dictionaries might be defined as result of operations and its combinations above anther dictionaries.

**Def. 5.5.2.1. Dictionary operations** are set theory operations $\cup, \cap, \times$, +, - for pair arguments, lexicographical and numerical orderings and several additional operations:

- **creating hierarchy of named** dictionaries**:** expression <dictionary A>(<dictionary B>,…,<dictionary C>); it defines a set of (empty) dictionaries with names A(b,…, c) for each combinations of elements of B,…, C; B,…,C are dictionaries or dictionary expressions with operations,

- **transformation serial dictionaries to another serial** dictionary: expression <operation> <hierarchical dictionaries>| <dictionary1>…| <dictionaryN>; here <operation> is one of symbols $\cup, \cap, \times$, +, - , <hierarchical dictionary> is a hierarchy of named dictionaries, dictionary1…, dictionaryN are one of argument in it expression; operation produces a several dictionaries with names $O(\cup A(B, c_1))$ ,…,$O(\cup A(B, c_m))$; each of them is a joint of all dictionaries of $A(b_i, c_p)$ for each $b_i \in B$ and fixed $c_p$,

- **expansion of hierarchical dictionary:** expression <hierarchical dictionary E>(<dictionary A>,…, <dictionary B>)//<dictionary C>//… define a dictionary of elements: $c_1$, followed with <joint of all dictionaries with any element of A,…, any element of B and fixed element $c_1$ from C>, then next element $c_2 \in C$ again follower with <joint of all dictionaries with any A,…, any element of B and fixed element $c_2$ from C>, the again c3,…..

Each elements of any dictionary keeps track of its origin via a value of special attribute called **genesis**. For example, element w of A⋃B is **genesis** has A, B if w belongs to A and B; if it belongs only to B, it is B only. These attribute is used for processing as well.

**Examples of operations**. Denote a PERSONAL list via P, a DEPARTMENT={IT, HR} via D, BRANCH={USA, Can, EMU} via B. The expression P(D, B) defines names for 6 personal lists: P(IT, USA), P(IT, Can), P(IT, EMU), P(HR, USA), P (HR, Can), P(HR, EMU).

The expression P (D, B)|D defines three dictionaries  -  one for each branches without dividing by departments:

Dictionary1:  name = P(USA) , elements={(Person1, **generics** USA∈B, IT∈D(USA)),  (Person2, **genesis** USA, IT∈D(USA)),  (Person4, **genesis** USA∈B, HR∈D(USA))};

Dictionary 2: name = P(Can), elements={ (Person3, **genesis** Can∈B,  IT∈D(Can))};

 Dictionary 3: name = P(EMU), elements={( Person5, **genesis** EMU∈B, HP∈D(EMU))}.

The expression P(D, B)//D gives a  plain dictionary:

   IT, **genesis** D(B)
   Person1, **genesis** USA∈B, IT∈D(USA)
   Person2, **genesis** USA∈B, IT∈D(USA)
  Person3, **genesis** Can∈B, IT∈D(Can)
   HR, **genesis** D(B)
   Person4, **genesis** USA∈B,HR∈D(USA)
   Person5, **genesis** EMU⋃B, HR∈D(EMU)

It consists of sections starting with department names and followed by a list of people from departments of each branches.

**Table series** are defined by types and four dictionaries for defining its structure and names:
  -   dictionary of names,
  -   dictionary of table attributes,
  -   dictionary of columns,
  -   dictionary of rows.

These dictionaries might be defined with expressions as well, but outside of tables.

Elements of table series dictionaries are **selected in parallel**, meaning if a dictionary name is in two places of expressions, then the same element will represent this list.
Each element of name dictionary determines a single table with rows and columns.

**Example 5.5.1.**  Let T be a table series:

name  = A
table attributes = U
rows =R(A, B)//B

columns = C(A).

Then a single table with name $a$ has columns from C($a$) and rows from list R($a$, B)//B. Generally, for different names the same type of table may have different columns and rows.

An elementary table object is a cell. It is identified by an element of name list, values of table attributes, element of row list and element of column list.

Some table series are a little bit unusual:
name: -
table attributes: -
rows : R
columns : C(R).

Here each row has its own set of columns.

More studies are needed to determine a more precisely meaning and sense of table series definitions.

As it was mentioned before, together with table processing applications it makes sense to have a collection of operations applicable to columns and rows from of the same or different table series. This set of operation defines all possible ways for table series processing and it may be considered as an operational semantics of data and possible ways for future transformations.  It also works for semantic errors detection in a future application evolution.

### 5.5.3. Table processing

**Def. 5.5.3.1. Table processing** is a mapping several table series, called **input tables**, to one table series, called **result table**. The mapping is based on dictionaries: the processing is allowed only for cells with consistent genesis when each dictionary from result tables cell's genesis is in genesis of input cells.

**Remark 5.5.0**. It make sense to have lists in description of output tables which are not in descriptions of these input tables and to keep its cells unchanged after this processing - they might be used for later processing or input.

**Example 5.5.3.1.**  Subject area for this example is determined by dictionaries. Company structure is determined via the following dictionaries:
BRANCHES = {br1,br2,br3}
and
DEP(BRANCHES) contains three dictionaries:

DEP(br1)={d11,d12,d13},
DEP(br2) ={d21,d22},
DEP (br3)={d31}.

Projects are determined with TYPES ={ r, s, h} (r-research, s-software, h-hardware), and departments DEP.

PROJECTS(DEP(BRANCH), TYPES)={PROJECTS(d11, r)={p1,p2}, PROJECTS(d12, r)={r3}, PROJECTS (d13, r)={r4},…, PROJECTS(d13,h)={r3,r4}} is lists of all projects for each BRANCH, DEPartment and project TYPES.

Input report tables are quarterly department reports:

type:  EXPENSE-REPORT
names : -
table attributes: DEP, BRANCH, QUARTER;
columns: INDICATORS;
rows: PROJECTS(DEP(BRANCH), TYPES));

where INDICATORS ={ expense, personal}.

The following is one of the tables of series EXPENSE-REPORT:

Table attributes

| Dep | Branch | Quart |
|-----|--------|-------|
| d21 | br2 | 3 |

| projects | expenses | personal |
|----------|----------|----------|
| p21 | 3 | 2 |
| p22 | 4 | 1 |
| P23 | 2 | 1 |
| P24 | 2 | 1 |

Genesis for a cell (p22, expenses) of this table is "PROJECTS p22, DEP d21, BRANCH br2, TYPE r, QUART 3", value of the cell is 4.

The instance of the output report table:

type:  **Expense-by-branches**
name**:-**
table attributes:-
column: INDICATORS,
rows: BRANCHES

is shown bellow:

| branches | expenses | personel |
|----------|----------|----------|
| br1 | 11 | 16 |
| br2 | 8 | 10 |
| br3 | 20 | 17 |

For cell (br1, expenses) of this table genesis is: "INDICATORS expenses, BRANCHES br2".

Collection of attributes for first table genesis includes the attributes of the second one. So there is a homomorphism of EXPENSE-BY-BRANCHES cells (department, branches, projects, expenses, personal) to EXPENSE-REPORT cells (expenses, personal, branches) and values of these cells can be process together.

For complete determination of the processing we need to add that to cells of columns 'expense' and 'personal' are applicable accumulative functions:

Expense-by-branches(expenses, branches) = **accum**(EXPENSE-REPORT(expense, branches),

Expense-by-branches(personal, branches) = **accum**(EXPENSE-REPORT(personal, branches).

Then all cell's values of homomorphism images are summarized to cells (expenses, branches) and (personal, branches) and we will get a final table of expenses and personal used in projects for each branches.

Adding "efficiency" column to table EXPENSE-BY-BRANCHES allows to obtain output table of type EXPENSE-EFFICIENCY with a column dictionary ADD-IND = INDICATORS ∪{efficiency}:

type: EXPENSE-EFFICIENCY
names: -
table attribute: -
columns: ADD-IND
rows: BRANCHES,

The computing formula is:
EXPENSE-EFFICIENCE(efficiency)=
EXPENSE-BY-BRANCHES (expenses) / EXPENSE-BY-BRANCHES(personal).

**REMARK 5.5.1**. Table series approach gives a formal base for defining important and illusive thing such as **full definition of subject matter**.

**REMARK 5.5.2**. Another important fact is that table series dictionaries largely define the **algorithm of its processing**. So program of it processing became smaller. Less qualification is needed for programming and debugging.

**REMARK 5.5.3.** For the analysis **unstructured text mining**, a reconstruction of a system of connected tables may be a **strategic target**, determining if and when we got all available info. Reconstructing a subject area – full set of tables without holes – is more important than collecting pieces of unknown size of information we got in one separate table.

This also determines an **intermediate targets** for mining: lists for table names, columns and rows, tables with common lists.

**Remark 5.5.4.** Many data have a temporary importance, and it is often hard and expensive to carry ballast data. Considering a business data as a flow of permanently coming and disappearing data (not all of it has to be accumulated) this description will help to build more adequate systems than one based on database concept.

Such **application** is a set of local table rules (LTR) and reserved (system) table LRRE of local rule names ready for execution. Rules have names and two collections: list of input tables and list of output tables. They are form a string in reserved table LTP-D with 3 columns: "LTR-name" for local table rule name, "LIT" for list of input tables and "LOT" for list of output tables. The final reserved table is LTP-In with three columns: "LTP-P name", "Initializing type" and "Initializing name".

The DDR Monitor starts when LRRE table is not empty. The LRRE table is updated if some ordinary application tables are updated or have been input by user or came from outside networks. Rule with empty initialized list also starts the application but never is executed again because input tables never updated - there are no ones. When started, the application never stops, it may only wait for updating some of input tables. After inputting or editing or creating table the Monitor checks the LTP-D table and adds these rows to LRRP. Then it executes rules from its rows with updated initialized tables.

If for some reason computed was restarted, the service again starts with LRRE. Because of any local rule does not change any application table - they might be updated only with the service - after restarting the computer, any unfinished steps will be repeated without any data corruption (as MSMQ does).

The DDR itself might functioning and growing the same way. For example, to add a distributed functionality it is enough to have a (system) rule for sending/receiving data to/from another nodes. To support securities, the only (system) rule to be initialized is a rule for checking authorization and outputting a LRRE.

This project might be good for student themes. It does not need any DB, complex OS and programming languages or any other expensive software. Working prototype might be finished in 3-5 months in LINUX and C++ or JAVA.

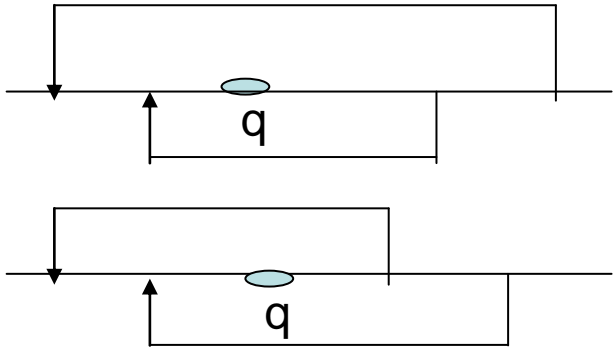

Figure 2.1.1. Structured and unstructured loops.

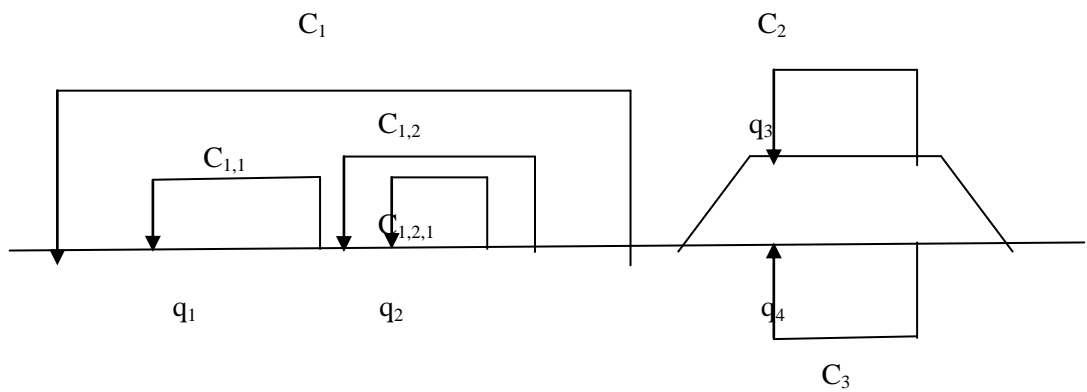

Figure 2.4.1. Loop naming

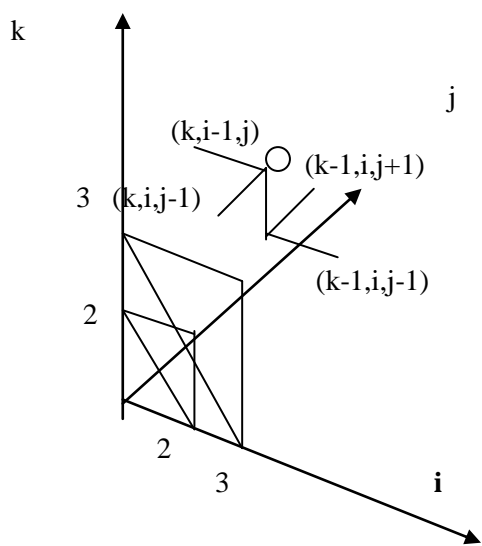

Figure 2.4.2. A hyper plains for parallel execution

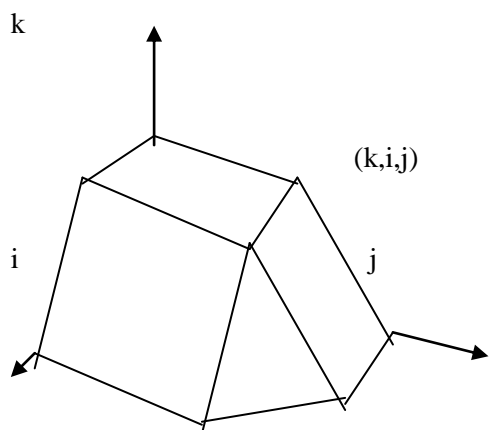

k

(k,i,j)

i

j

Figure 2.4.3. Cone for point (k,i,j)

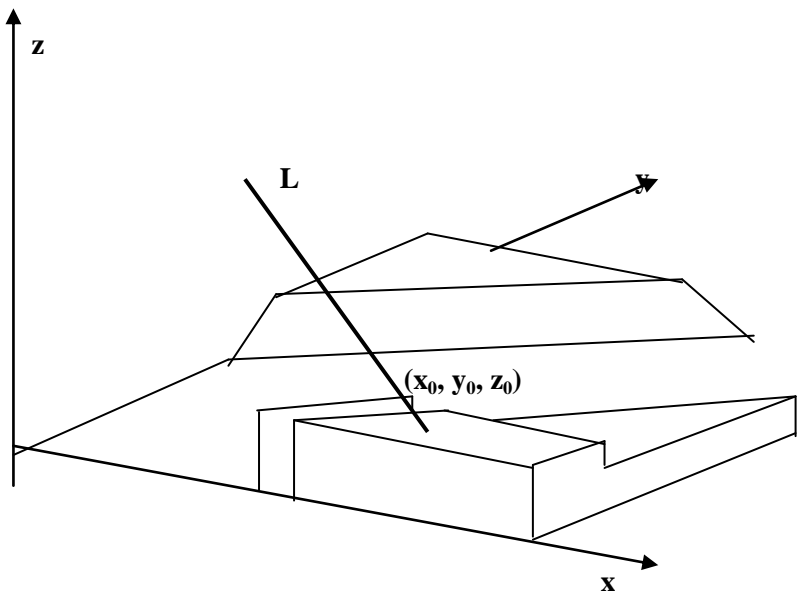

Figure 2.4.4. Cone of dependence for system from an example 2.4.2.

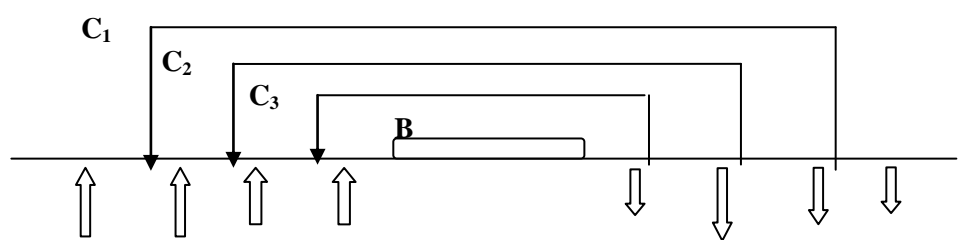

Figure 3.1.1. Nested loops

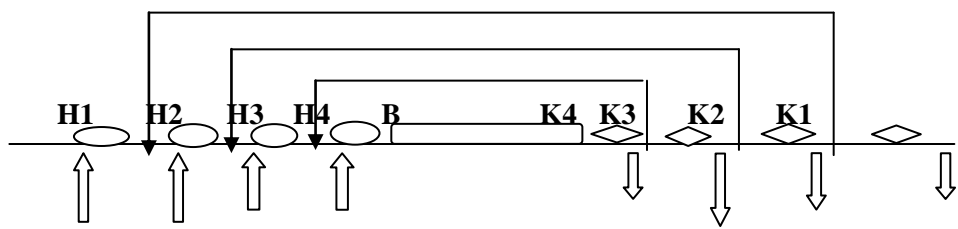

Figure 3.1.2. Nested loops with starter and final body parts.

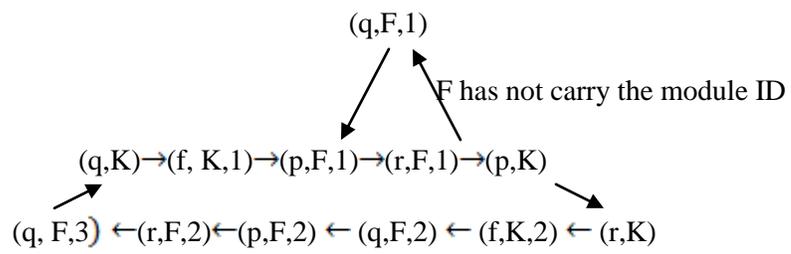

Figure 3.2.1. A transition graph of a control automata for data exchange